# Lithium-ion battery degradation: using degradation mode analysis to validate lifetime prediction modelling

Ruihe Li[1,2], Niall D. Kirkaldy[1,2,†], Fabian Oehler[3], Monica Marinescu[1,2], Gregory J. Offer[1,2], Simon E. J. O'Kane[1,2]

[1]Department of Mechanical Engineering, Imperial College London, UK

[2]The Faraday Institution, UK

[3]Chair of Electrical Energy Storage Technology, Technical University of Munich, Germany

[†]Current address: Elysia Battery Intelligence, Oxford, UK

## Abstract

Predicting lithium-ion battery lifetime is one of the greatest unsolved problems in battery research right now. Recent years have witnessed a surge in lifetime prediction papers using physics-based, empirical, or data-driven models, most of which have been validated against the remaining capacity (capacity fade) and sometimes resistance (power fade). However, there are many different combinations of degradation mechanisms in lithium-ion batteries that can result in the same patterns of capacity and power fade, making it impossible to find a unique validated solution. Experimentally, degradation mode analysis involving measuring the loss of lithium inventory, loss of active material at both electrodes, and electrode drift/slippage has emerged as a state-of-the-art requirement for cell degradation studies. In this paper we coupled five degradation mechanisms together for the first time. We also showed how three models with different levels of complexity can all fit the remaining capacity and resistance well, but only the model with five coupled degradation mechanisms could also fit the degradation modes at all temperatures. This work proves that validating only against capacity and power fade is no longer sufficient, and state-of-the-art experimental and modelling degradation studies should include degradation mode analysis for validation in the future.

# Introduction

Due to the requirements in electric vehicles, smart phone and energy storage stations, the demand of lithium-ion batteries (LIBs) is expected to increase by 33% each year from 2022 to reach ~4700GWh by 2030[1]. The performance of LIBs degrades with time and repeated cycling[2]. The production and recycling of LIBs poses huge environmental and financial challenges to the whole society[3], making degradation of LIBs a big concern.

To understand the degradation behaviours of LIBs, a computational model is required. Among different types of models, physics-based models are useful because they account for the root causes of degradation. Examples of the usage of physics-based degradation models include predicting remaining useful life (RUL)[4], optimising operation conditions[5] and improving manufacturing procedures[6]. The insights from physics-based models can also feed into empirical models and data-driven models to get more flexibility and reduce computational time[7].

To achieve the above benefits, the model must be well-parameterized and validated against experimental measurements. The gold standard for the last decade has been to reproduce multiple ageing features, normally in the form of capacity retention curves, under as many ageing conditions as possible. For example, the square root of time dependency can be reproduced by a diffusion limited process[8]. The temperature dependency can be depicted by two Arrhenius relationships: one for high temperatures[8, 9], and one for low temperatures, with an optimum operating temperature in the middle around 25°C[10, 11]. Capacity recovery during early stage of ageing can be captured by considering anode overhang[12]. Rollover failure can be explained by SEI coupling with lithium plating[13], or particle cracking coupling with SEI on cracks[11], or SEI cracking coupling with electrode dry-out[14].

In recent years, researchers have realized the importance of coupling different ageing mechanisms together[11, 15-17]. However, the number of fitting parameters in these coupled models is already exceeding 10. Many of them are not yet possible to measure with classical electrochemical tests. Instead, they can only be obtained through directly fitting the ageing data. In that case, overfitting becomes a big concern. Most previous papers have validated their models against capacity retention, which is the simplest and most easily measured performance index. There are a few that validate against resistance[18, 19], full cell dQ/dV[20, 21], and even SEI thickness[22]. More recently, degradation mode (DM) analysis has been used to provide additional information linking degradation mechanisms and degradation effects[23]. This work focuses on three DMs: loss of lithium inventory (LLI), loss of active material in the negative electrode ($LAM_{NE}$) and in the positive electrode ($LAM_{PE}$). Baure and Dubarry found that the $LAM_{NE}$:LLI ratio can be an effective index to identify accelerated degradation[24]. Therefore, DMs have been used as both parametrization and validation indices in empirical ageing models[23]. However, to our best knowledge, DMs have never been used as validation indices in physics-based models. In this work, we highlight the importance of DMs by showing that, for a simple degradation study, three models can fit capacity and resistance equally well, but only two of them also fit LAM, and only one can fit the data at all temperatures reasonably well.

**Ageing mechanisms**

The degradation model used here is based on that of O'Kane *et al.*[11], with the addition of Li *et al.*'s [25] model of solvent consumption and electrolyte drying. Fig. 1 illustrates the five different mechanisms and how they interact with each other.

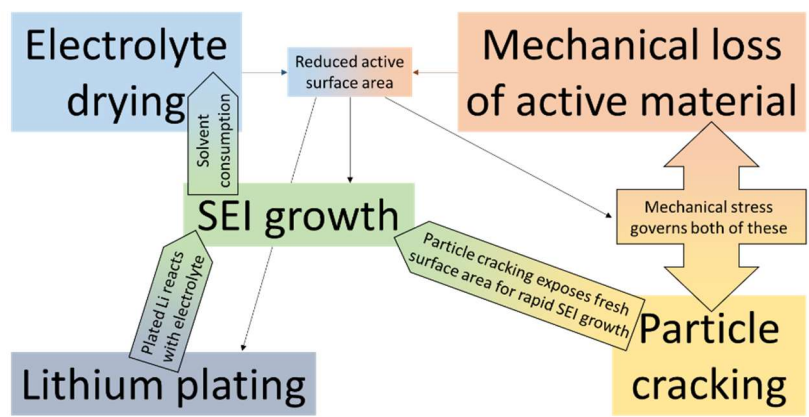

Fig. 1 The 5 degradation mechanisms in our model and how they interact (adapted from O'Kane *et al.* [11]).

To start with, SEI layer growth will consume useable lithium inventory and solvent, potentially leading to electrolyte drying. The electrolyte drying will then make part of the active surface area inaccessible. Mechanical loss of active material will also reduce the active surface area. Under the same applied current (to the cell), the remaining active surface area will have to sustain higher interfacial current density, accelerating SEI growth, particle cracking, and mechanical induced LAM. Particle cracking exposes fresh surface area to the electrolyte, which triggers rapid SEI layer growth. The increased interfacial current density also accelerates lithium plating. The plated lithium can then react with electrolyte and form SEI.

However, note that all these mechanisms may not show the same significance in one LIB cell under one specific ageing condition. Therefore, we may not need all of them to fit one specific set of experimental data. However, SEI layer growth is the most common side reaction and used in almost all previous physics-based modelling papers. Solvent consumption is a side effect of SEI layer growth. Therefore, in this work, we have picked three different combinations of sub-models as examples of our ageing models: (1) *SEI only*; (2) SEI with solvent consumption, which we call *SEI + Dry out*; (3) all mechanisms in Fig. 1 included, which we call *5 coupled*. We will show the results of best fit of all these three models against: (1) the commonly used indices, voltage, capacity retention, and resistance; (2) the indices that have been omitted, i.e., the degradation modes (DMs).

**Model validated against voltage, capacity, and resistance**

We first validate our ageing model with reference performance tests (RPTs). The detailed protocols of RPTs and model settings can be found in Kirkaldy et al.[26] and Supplementary Information, respectively. Fig. 2 shows the C/10 discharge voltage of the experiment and the three models aged at 25°C at beginning of life (BOL), middle of life (MOL), and end of life (EOL). In Fig. 2, the three models show excellent agreement with each other and the experimental data. The excellent agreement is further confirmed by the mean percentage errors (MPEs, Fig. S2 and Table 1) and root mean square error (RMSE, Fig. S3 and Table S8). To be specific, the average MPEs of all RPTs are all below 0.8% (Table 1) and the average RMSE are all below 40 mV for the 3 models, indicating very good fits for the voltage. Therefore, the 3 models all perform well for voltage validation. The *5 coupled* model is slightly better than the *SEI + Dry out* model; and the *SEI + Dry out* model is slightly better than the *SEI only* model.

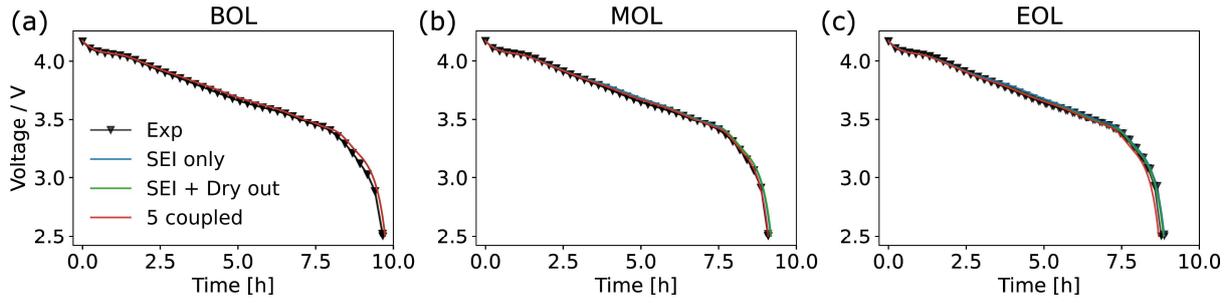

Fig. 2 Comparison of the three models during C/10 discharge against experimental data at 25℃: (left) beginning of life (BOL); (middle) middle of life (MOL); (right) end of life (EOL).

Table 1. Average MPE of voltage fitting for all C/10 discharge RPT cycles.

| Average MPE / % | 10°C | 25°C | 40°C |
|---|---|---|---|
| *SEI only* | 0.72 | 0.70 | 0.80 |
| *SEI + Dry out* | 0.58 | 0.58 | 0.63 |
| *5 coupled* | 0.44 | 0.48 | 0.52 |

Fig. 3 shows the RPT results for state of health (SOH, upper column) and lump resistance (lower column) for three temperatures. SOH is defined as the C/10 capacity during each RPT over that of the first RPT. The lump resistance is obtained by the voltage drop 0.1 s after the 12$^{th}$ pulse of a C/2 GITT discharge, see Experiment details (RPT and ageing) in SI. The two light grey lines are the result of the two tested cells, and the black line represents their average. The three coloured lines correspond to the three models. The MPEs of SOH and lump resistance of the three models are calculated in Table 2. Based on Fig. 3 and Table 2, the *SEI only* and *SEI + Dry out* model fit SOH at 25°C pretty well, with MPEs of 0.15 and 0.32, respectively. However, both overestimate SOH at 10°C and 40°C. The *5 coupled* model, by contrast, has better fits of SOH at 10°C and 40°C, with MPEs of 0.42 and 0.57, but underestimate SOH at 25°C. The lump resistance predicted by the three models has the same trend as the experiments for 25 °C, but the resistances predicted for 10 °C and 40 °C are too high. Overall, all three models fit SOH well, with a maximum MPE less than 0.88%. The fitting results on the lump resistance are less accurate. However, for MPEs of resistance, the three models again perform similarly. There are no models that perform much better than the others based on the fitting results of SOH and lump resistance.

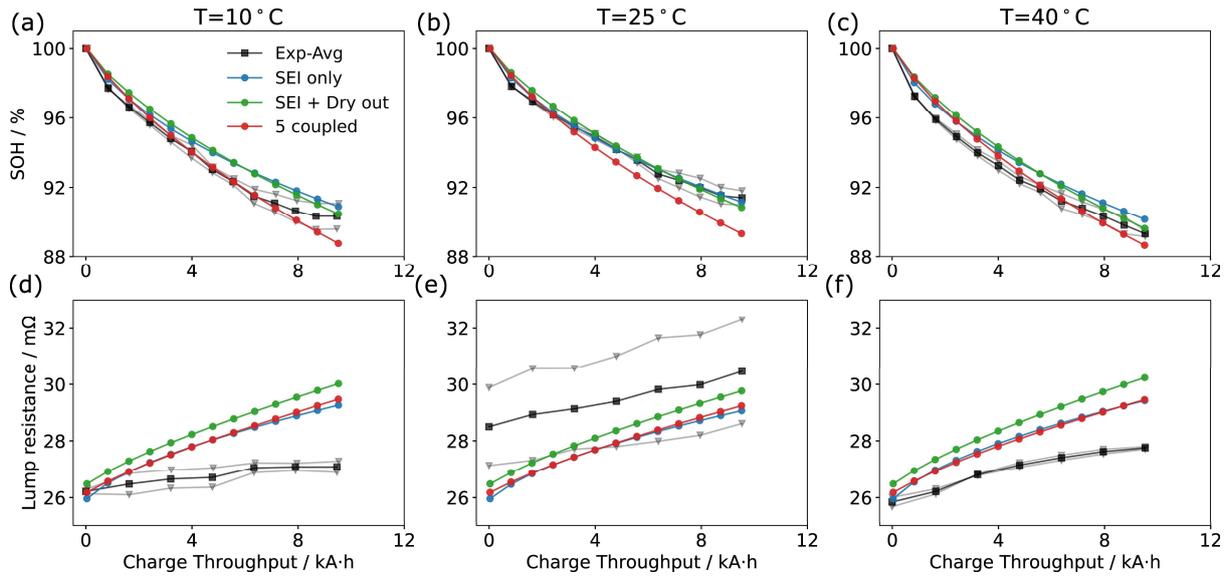

Fig. 3 SOH and lump resistance of the 3 models and experiment.

## Model validated against degradation modes (DMs)

To further compare the 3 models, we validate them against DMs (Fig. 4). As presented in Fig. 4, all three models fit LLI with different accuracies. However, the *SEI only* model has zero LAM in both electrodes, which leads to 100% MPEs. The other two models have included LAM in their formulae and therefore both have LAM in the two electrodes.

For all three temperatures, the *SEI only* model significantly underestimates LLI and therefore has the highest MPEs (Table 2). However, this simple model still achieves an excellent fit to SOH. These findings appear to contradict each other. However, recall that the SOH is evaluated using a C/10 discharge as opposed to a true OCV. The resistance therefore plays a role. The higher the resistance, the sooner the C/10 discharge reaches the lower voltage cut-off. It is therefore possible for the *SEI only* model to predict the same SOH as the other two models even if LLI and LAM in both electrodes are all lower, because the resistance is higher.

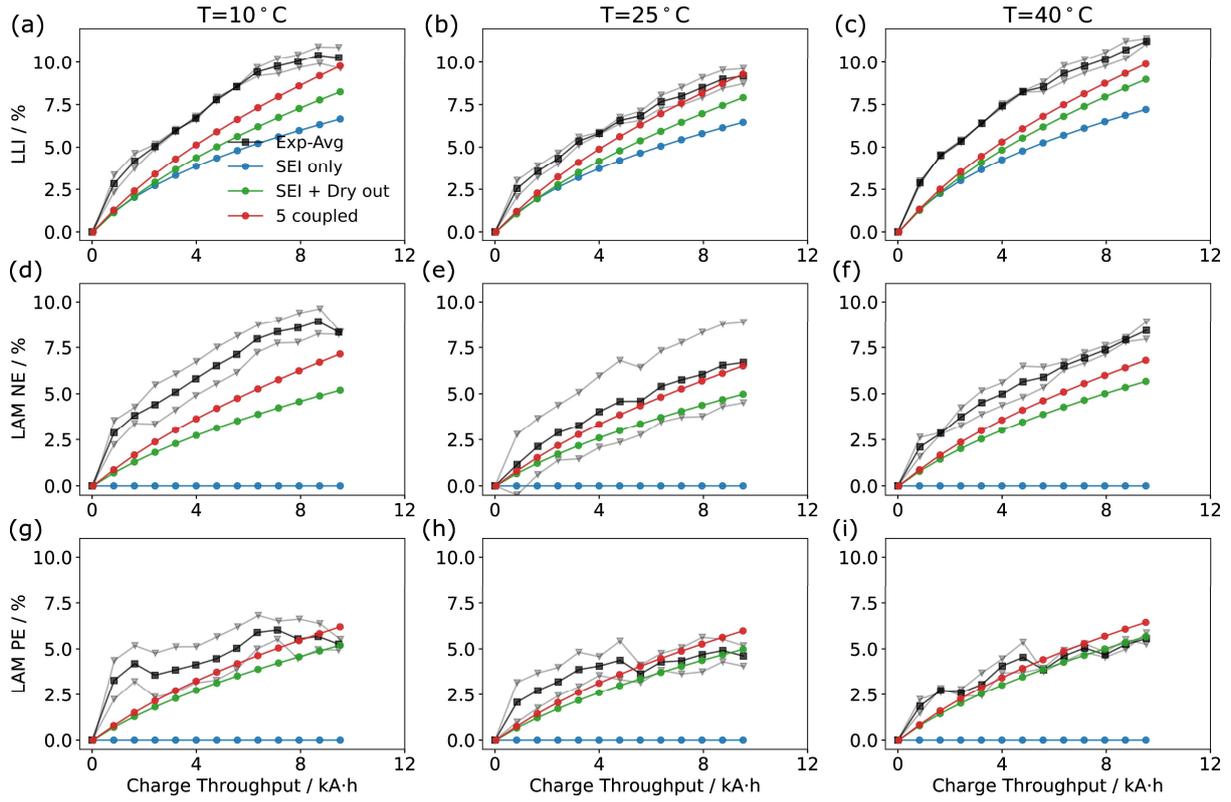

Fig. 4 Degradation mode analysis of the 3 models and experiment.

The *SEI + Dry out* and *5 coupled* models fit all DMs well at 25 °C, with MPEs under 54%. (Table 2). The *5 coupled* models fit all DMs better than the *SEI + Dry out* model at all three temperatures, highlighting the importance of including detailed degradation mechanisms. Now that we have presented 5 different indices to evaluate the degradation model, it is necessary to get a weighted index based on them to assist the overall evaluation. However, there are no well-acknowledged weighting methods on this. Therefore, we propose the following weighting ratio:

$$\text{MPE}_{tot} = \frac{1}{2}\text{MPE}_{SOH} + \frac{1}{8}\text{MPE}_{Res} + \frac{1}{8}\text{MPE}_{LLI} + \frac{1}{8}\text{MPE}_{LAM\_NE} + \frac{1}{8}\text{MPE}_{LAM\_PE}. \qquad (1)$$

We have given greater weighting to SOH as it is the most highly desired property for end users. The result of $\text{MPE}_{tot}$ is listed in Table 2. The *5 coupled* model has lower $\text{MPE}_{tot}$ (11.99%+10.41%+12.05=34.45%) for the three temperatures compared to that of the *SEI + Dry out* model (16.91%+13.78%+14.45%=45.14%). The *SEI only* model has the highest $\text{MPE}_{tot}$ due to the 100% MPEs in LAM for both electrodes. The performance of the three models is initially difficult to be distinguished under voltage, SOH, and lump resistance but now quite clear using DMs.

Table 2. Mean percentage error for all degradation modes, models, and temperatures. The "Total" column is a weighted index defined in Eq. (1).

| Model | T / °C | SOH | Res | LLI | LAM$_{NE}$ | LAM$_{PE}$ | Total |
| --- | --- | --- | --- | --- | --- | --- | --- |
| SEI only | 10 | 0.83 | 4.41 | 44.45 | 100.00 | 100.00 | 31.52 |
| SEI only | 25 | 0.15 | 5.74 | 41.69 | 100.00 | 100.00 | 31.00 |
| SEI only | 40 | 0.86 | 3.77 | 45.86 | 100.00 | 100.00 | 31.63 |
| SEI + Dry out | 10 | 0.87 | 6.13 | 36.25 | 53.57 | 35.88 | 16.91 |
| SEI + Dry out | 25 | 0.32 | 4.05 | 33.84 | 38.17 | 32.89 | 13.78 |
| SEI + Dry out | 40 | 0.86 | 5.77 | 37.56 | 44.36 | 24.45 | 14.45 |
| 5 coupled | 10 | 0.42 | 4.49 | 25.10 | 37.64 | 27.02 | 11.99 |
| 5 coupled | 25 | 0.88 | 5.50 | 22.75 | 20.58 | 30.94 | 10.41 |
| 5 coupled | 40 | 0.57 | 3.68 | 31.58 | 34.18 | 24.64 | 12.05 |

There is a direct relationship between the DMs and the changes in the half-cell potential curves of the negative and positive electrodes. This is illustrated in Fig. 5 for C/10 constant current discharges of the cell aged at 10 °C. results for 25 °C and 40 °C are presented in Fig. S9 and Fig. S10, respectively. Comparing the simulations at BOL and EOL, the *SEI only* model leads to a shift of the half-cell potential curves in relation to each other, which can be recognised in the differential voltage analysis (DVA) in Fig. 5 (c) by the leftward shift of the positive electrode potential. The capacity fade therefore results directly from the shift in electrode balancing caused by LLI.

For *SEI + Dry out*, LAM in both electrodes caused by dry out shows up as lower cell voltage during discharge. The influence of LAM$_{NE}$ is more significant due to the steeper slope of the open-circuit potential (OCP) curve of the negative electrode compared to the positive electrode. Fig. 5 (c) shows LAM$_{PE}$ as a compression of the respective DVA compared to initial conditions. Considering the additional DMs summarised in Fig. 1, the *5 coupled* model estimates a considerably higher capacity fade at 10 °C (cf. Fig. 3). The LAM-induced LLI is visible as strongly shifted positive electrode potential curve in Fig. 5 (c) and the corresponding DVA.

**Ruihe observation:**

1. full cell DVA: (1) aged *SEI only* on top of BOL at the 1.2V peak; (2) aged *SEI + Dry out* and aged *Full* shifted left at the 1.2V peak; → these originate from Neg DVA; (3) the other part of the 3 aged curves shifted left compared with BOL; (4) curves shrink overall.

2. Neg DVA: (1) aged *SEI only* on top of BOL at the 1.2V peak; (2) aged *SEI + Dry out* and aged *Full* shifted left at the 1.2V peak; (3) the other part of the 3 aged curves shifted left compared with BOL; (4) curves shrink overall.

3. Pos DVA: all 3 aged curves shifted to left .

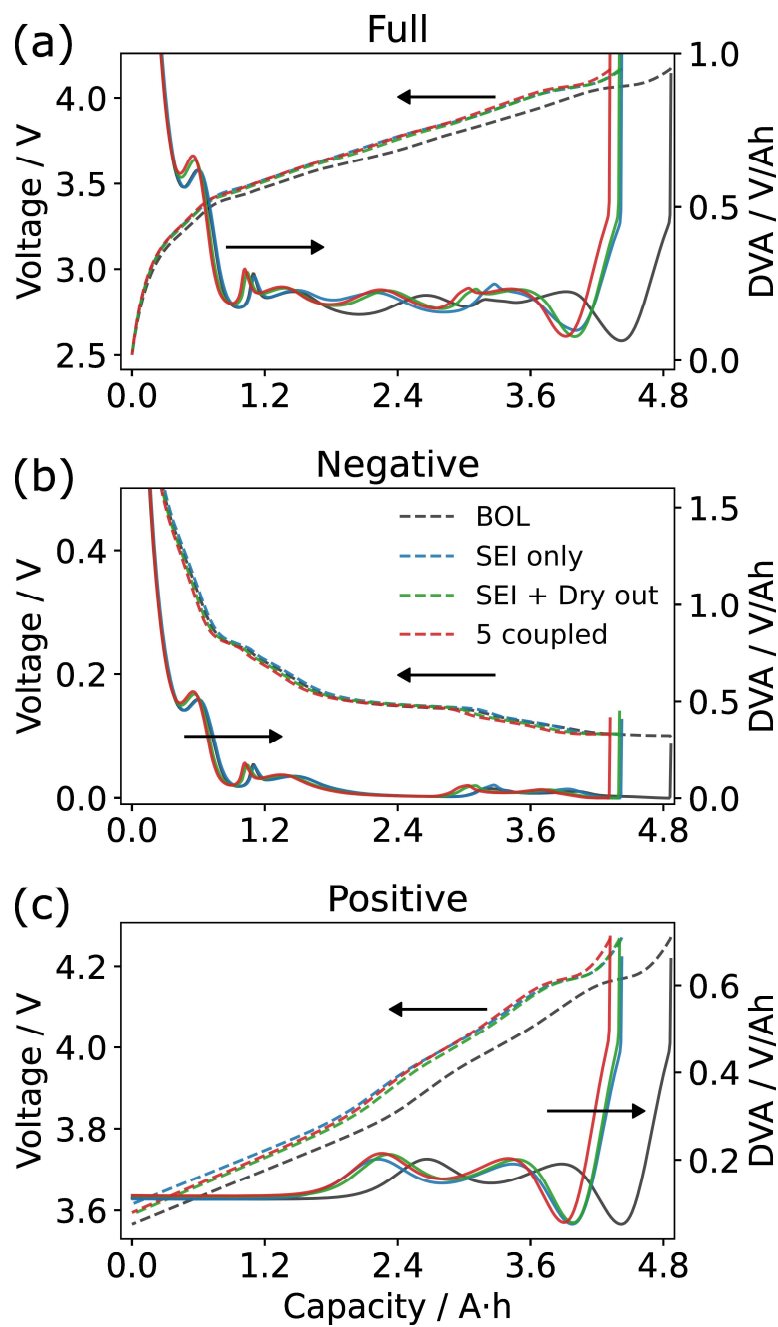

Fig. 5 Effect on half-cell potentials (dashed lines) and DVA (solid lines) during C/10 charge of the cell aged at 10 °C (BOL vs. EOL-RPT-13).

## Discussion / Conclusion

Any truly predictive degradation model of LIBs must be able to capture all of the following: (1) degradation modes; (2) different degree of LAMs in two electrodes; (3) temperature dependence. Firstly, for cells showing degradation in not just LLI but also LAM, the model must include mechanisms of LAM. Secondly, though the dry-out sub-model includes LAM, it causes the same amount of LAM in two electrodes, which is not usually what is measured. That calls for additional mechanisms which induce different LAM in the two electrodes. Thirdly, the model needs to reproduce the temperature dependence of the experimental data.

Although the first standard excludes the *SEI only* model, and the second standard excludes the *SEI + Dry out* model, we find that the temperature dependencies can be reproduced by all three models. One of the well-known temperature dependencies is that cells degrade more quickly at low temperatures due to lithium plating and particle cracking, and at high temperatures due to SEI growth, compared to medium temperatures at which they degrade more slowly. For the cells we study here, the total capacity fade is most severe under 40 °C, followed by 10 °C and 25 °C. The capacity loss vs temperature curve thus exhibits a "V" shape with the 40 °C end being higher, see Fig. S8 (d). The main parameters affecting the temperature dependency of the *SEI only* and *SEI + Dry out* models are the SEI growth activation energy ($E_{\text{act}}^{\text{SEI}}$) and the negative electrode diffusivity activation energy $E_{\text{act}}^{D_{s,n}}$). The electrochemical reaction that forms SEI has thermally activated kinetics and is therefore assumed to follow an Arrhenius relationship with activation energy $E_{\text{act}}^{\text{SEI}}$. However, Li diffusion in the electrode particles is also thermally activated and follows its own Arrhenius relationship with activation energy $E_{\text{act}}^{D_{s,n}}$. The lower the diffusivity, the lower the negative electrode potential during charge, increasing the SEI growth rate. These competing effects make the temperature dependence of SEI growth more complex than is commonly assumed in the literature. To investigate such competing effects, we have chosen three values of $E_{\text{act}}^{\text{SEI}}$ (5e2, 5e3, 5e4 J/mol) and $E_{\text{act}}^{D_{s,n}}$ (2e4, 4e4, 6e4 J/mol), respectively, making 9 combinations. The resulting temperature dependencies are presented in Fig. S8.

For the *SEI only* model, three combinations of ($E_{act}^{SEI}$, $E_{act}^{D_{s,n}}$) can reproduce the experimented observed temperature dependency (Fig. S8 (a)), namely (5e3, 2e4) J/mol, (5e3, 4e4) J/mol, (1e4, 6e4) J/mol. For the *SEI + Dry out* model, LLI has two sources

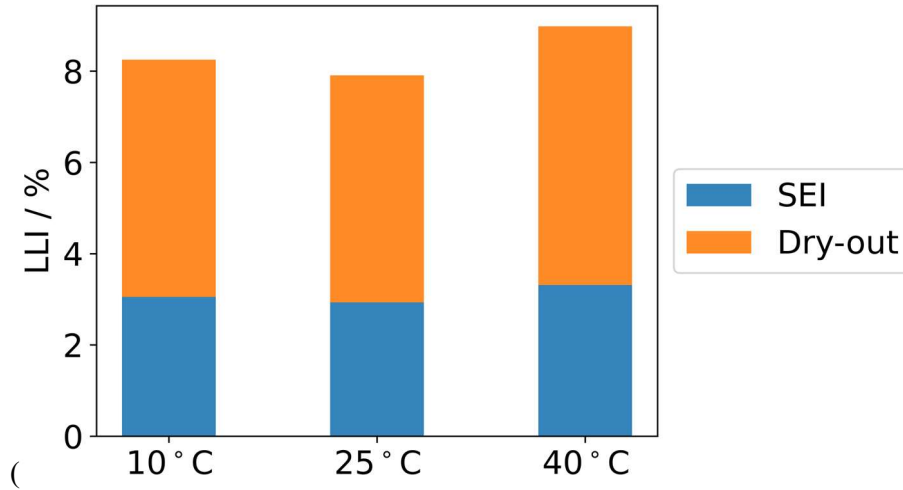

(Fig. S5): lithium locked in the SEI and lithium trapped electrode particles that have become inactive due to dry-out induced LAM. As presented in LLI of the *5 coupled* model has 4 sources: SEI, SEI on cracks, lithium plating and LAM, among which LAM accounts for the largest portion (Fig. S6). The LAM induced LLI can be further decomposed into 2 parts: the dry-out induced LLI and the stress-driven LAM induced LLI. As presented in Fig. S7, the temperature dependence of both LLI and LAM are mainly caused by the dry-out induced LLI. However, the temperature dependency of stress-driven LAM relies only on $E_{act}^{D_{s,n}}$, i.e., a higher value of $E_{act}^{D_{s,n}}$ gives a lower solid diffusivity under low temperatures and therefore the higher stress-driven LAM. As a result, only two combinations of $(E_{act}^{SEI}, E_{act}^{D_{s,n}})$ can reproduce the desirable temperature dependency, namely (5e3, 2e4) J/mol and (5e3, 4e4) J/mol.

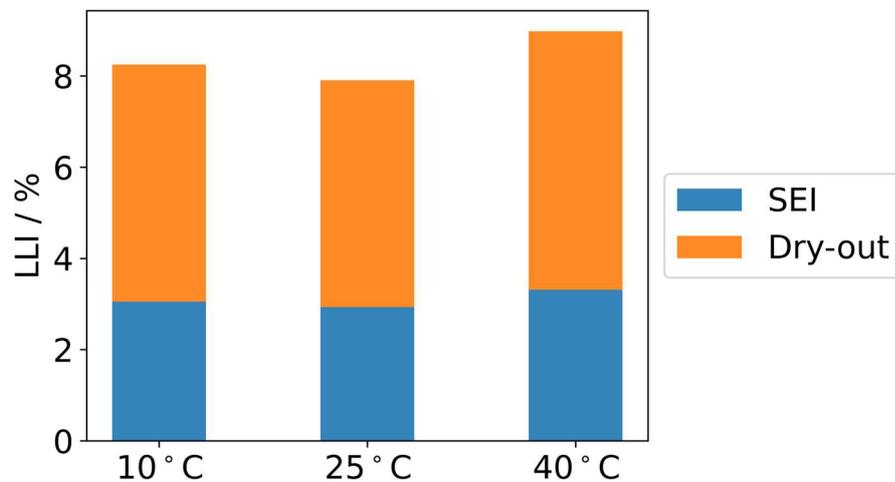

Fig. S5, LLI due to dry-out is almost twice that due to SEI under all three temperatures. However, dry out is a direct result of the SEI growth. Therefore, the total LLI and total capacity loss of the *SEI + Dry out* model follows the same temperature dependency (Fig. S8 (b)) as that of the *SEI only* model.

Despite the large parameter space of the model and the resources required to run each simulation (4GB of memory, 2 CPUs, and 16 hours on average for the *5 coupled* model), we were able to achieve the fits shown here using brute force. In the future, a thorough sensitivity study and optimization will be carried out to get a better fit and further explore the predictive power of the model.

**Methods**

As in O'Kane *et al.*, [11] the Doyle-Fuller-Newman (DFN) pseudo-2D model of LIBs is chosen for representing the beginning of life behaviour of the battery. Updated parameters for the LG M50 have since been published by O'Regan *et al.*, [27] which allow us to add a lumped thermal model and concentration-dependent diffusivities in the electrode particles, neither of which were included in O'Kane *et al.*'s [11] model. Dr. O'Regan also performed updated measurements of the half-cell open-circuit potentials (OCPs), which are consistent with their group's earlier measurements from Chen *et al.* [28] but extend over larger lithiation ranges, improving the accuracy of the degradation mode analysis. Details of these changes can be found in Zero-order hysteresis *model*

PyBaMM has an optional zero-order hysteresis model, in which the either the lithiation or delithiation OCP is used, depending on the sign of the current:

$$U_p^{OCP} = \frac{1 + \tanh[100(J + 0.2)]}{2} U_{p,lith}^{OCP} + \frac{1 - \tanh[100(J + 0.2)]}{2} U_{p,delith}^{OCP} \tag{S28}$$

$$U_n^{OCP} = \frac{1 - \tanh[100(J + 0.2)]}{2} U_{n,lith}^{OCP} + \frac{1 + \tanh[100(J + 0.2)]}{2} U_{n,delith}^{OCP} \tag{S29}$$

where the applied current density J has units of A/m2, is positive for discharge and negative for charge. The offset of 0.2 A/m2 is added so that the OCP corresponding to discharge is used when the cell is at rest. If it was not added, the hysteresis model would interfere with the GITT characterization because the OCP would change during the rest phases. The value of 0.2 A/m2 is chosen because the magnitude of the current density curing the CV charge is always greater than this, so the hysteresis model will not interfere with the CV charge either.

Input parameters in SI.

**Ageing sub-models**

*SEI growth*

O'Kane et al. chose a simple solvent-diffusion limited model of SEI growth, on the grounds that it was able to predict the square root of time dependence observed throughout the literature while only needing one adjustable parameter. The problem with the solvent-diffusion limited model is that it has no dependence on SoC, despite SoC dependence also being observed throughout the literature. In this work, we follow the assumption made by Single et al.[29] and von Kolzenberg et al.[30] that the SEI reaction rate is limited by the diffusion of the lithium interstitial within the inner SEI layer, leading to an expression of interfacial SEI current density of:

$$j_+^{\text{SEI}} = \frac{c_{\text{int,Li}}}{L_{\text{SEI}}^{\text{inner}}} \cdot D_{\text{int}} F \cdot \exp(-\phi_s + \phi_e), \tag{2}$$

among which $L_{\text{SEI}}^{\text{inner}}$ is the thickness of the inner SEI layer. $c_{\text{int}}$ and $D_{\text{int}}$ are the concentration and diffusivity of the lithium interstitial in the SEI layer, respectively.

For simplicity, we also assume that the inner and outer SEI layers grow at the same rate:

$$j_+^{\text{SEI,inner}} = \alpha j_+^{\text{SEI}}, \tag{3}$$

$$j_+^{\text{SEI,outer}} = (1-\alpha) j_+^{\text{SEI}}, \tag{4}$$

$\alpha$ is the inner SEI reaction proportion and set to 0.5 in this study.

The temperature-dependent SEI current density is:

$$j_+^{\text{SEI}}(T) = j_+^{\text{SEI}}(T_{\text{ref}}) \cdot \exp\left(\frac{E_{\text{SEI}}}{RT_{\text{ref}}} - \frac{E_{\text{SEI}}}{RT}\right), \tag{5}$$

where $E_{\text{SEI}}$ is the activation energy of the SEI reaction, $T_{\text{ref}}$ is the reference temperature, which is set to 298.15 K in this study (25°C).

The SEI thickness increases as followed:

$$\frac{dL_{\text{SEI,inner}}}{dt} = \frac{j_+^{\text{SEI,inner}}}{2F} \cdot \bar{V}_{\text{SEI}}^{\text{inner}} = \frac{c_{\text{int}}}{2L_{\text{SEI}}^{\text{inner}}} \cdot D_{\text{int}} \bar{V}_{\text{SEI}}^{\text{inner}} \cdot e^{-(\phi_s - \phi_e)} \qquad (6)$$

$$\frac{dL_{\text{SEI,outer}}}{dt} = \frac{j_+^{\text{SEI,outer}}}{2F} \cdot \bar{V}_{\text{SEI}}^{\text{outer}} = \frac{c_{\text{int}}}{2L_{\text{SEI}}^{\text{outer}}} \cdot D_{\text{int}} \bar{V}_{\text{SEI}}^{\text{outer}} \cdot e^{-(\phi_s - \phi_e)} \qquad (7)$$

where $F$ and $a_n$ are the Faraday constant and the specific surface area, respectively. For spherical particles $a_n = 3\varepsilon_n/R_n$, where $R_n$ is the radius of the negative electrode particles. $\bar{V}_{\text{SEI}}^{\text{inner}}$ and $\bar{V}_{\text{SEI}}^{\text{outer}}$ are the partial molar volume of the inner and outer SEI layers, respectively.

The total SEI thickness is the summation of the inner and outer SEI thicknesses:

$$L_{\text{SEI}} = L_{\text{SEI,inner}} + L_{\text{SEI,outer}} \qquad (8)$$

The SEI has an Ohmic resistivity $\rho_{\text{SEI}}$, which results in an overpotential $\eta_{\text{SEI}}$:

$$\eta_{\text{SEI}} = \rho_{\text{SEI}} \cdot j_n^{\text{tot}} \cdot L_{\text{SEI}}, \qquad (9)$$

where $j_n^{\text{tot}}$ is the total interfacial current density in the negative electrode.

This interstitial-diffusion limited model is an improvement on the solvent-diffusion limited model chosen because it captures both time dependence and SoC dependence despite still having only two adjustable parameters, $c_{\text{int}}$ and $D_{\text{int}}$, which effectively act as one parameter because they appear in (6) and (7) as a product of each other, never on their own.

*Lithium plating*

The lithium plating model in this study is unchanged from O'Kane et al.[11], who used a partially reversible plating model in which plating, stripping and dead lithium formation occur at the same time. The plating and stripping reactions are governed by a Butler-Volmer equation:

$$j_{\text{Li}} = Fk_{\text{Li}} \left( c_{\text{Li}} \exp\left(\frac{F\alpha_{a,\text{Li}}(\phi_s - \phi_e - \eta_{\text{SEI}})}{RT}\right) - c_e \exp\left(-\frac{F\alpha_{c,\text{Li}}(\phi_s - \phi_e - \eta_{\text{SEI}})}{RT}\right) \right), \qquad (10)$$

among which $j_{\text{Li}}$ is the  $k_{\text{Li}}$ is the lithium plating kinetic rate constant (in m/s), $c_{\text{Li}}$ is the concentration of the plated lithium, $\alpha_{a,\text{Li}}$ and $\alpha_{c,\text{Li}}$ are plating and stripping transfer coefficients respectively, which satisfy $\alpha_{a,\text{Li}} + \alpha_{c,\text{Li}} = 1$. The differential equation for $c_{\text{Li}}$ is

$$\frac{\partial c_{\text{Li}}}{\partial t} = -\frac{a_n j_{\text{Li}}}{F} - \frac{\partial c_{\text{dl}}}{\partial t}, \tag{11}$$

where $c_{\text{dl}}$ is the concentration of the dead lithium with its own differential equation

$$\frac{\partial c_{\text{dl}}}{\partial t} = \gamma c_{\text{Li}}. \tag{12}$$

$\gamma$ is a decay rate, defined as:

$$\gamma = \gamma_0 \cdot \frac{L_{\text{SEI},0}}{L_{\text{SEI}}}, \tag{13}$$

where $\gamma_0$ is the decay rate constant, a fitting parameter, and $L_{\text{SEI},0}$ is the initial thickness of the SEI layer. The dependence on $L_{\text{SEI}}$ is designed to account for the role played by solvent molecules from the electrolyte in the transition from plated lithium to dead lithium.

*Cracking*

The particle cracking is induced by cyclic stress. Therefore, we need to introduce the classic stress model first. This stress model is originally proposed by Zhang et al.[31], based on the equilibrium of stresses for a free-standing spherical electrode particle. The analytical solutions for the radial stress $\sigma_r$, tangential stress $\sigma_t$ and displacement $u$ are:

$$\sigma_r = \frac{2\Omega E}{(1-v)} \cdot [c_{\text{avg}}(R_i) - c_{\text{avg}}(r)], \tag{14}$$

$$\sigma_t = \frac{\Omega E}{(1-v)} \cdot [2c_{\text{avg}}(R_i) + c_{\text{avg}}(r) - \bar{c}/3], \tag{15}$$

$$u = \frac{(1+v)}{(1-v)} \cdot \Omega r c_{\text{avg}}(r) + \frac{2(1-2v)}{(1-v)} \Omega r c_{\text{avg}}(R_i), \tag{16}$$

where $\Omega$ is the partial molar volume, $E$ is the Young's modulus, $v$ is the Possion's ratio, $R_i$ is the radius of the particle and $c_{\text{avg}}(r)$ is the average Li$^+$ concentration between 0 and $r$:

$$c_{\text{avg}}(r) = \frac{1}{3r^3} \int_0^r \bar{c} r^2 \text{d}r, \tag{17}$$

where $\bar{c} = c - c_{\text{ref}}$ is the deviation in lithium concentration from the reference value $c_{\text{ref}}$ for the stress-free case.

Deshpande et al.[32] assumes that the tensile tangential stress ($\sigma_t > 0$) induces identical micro cracks on the electrode particle surface. Three parameters are proposed to describe these cracks, namely the length

$l_{cr}$, width $w_{cr}$, and density (number of cracks per unit electrode surface area) $\rho_{cr}$. It is further assumed that these cracks grow in length during cycling but maintain the same width and density. The growth of the crack length follows Paris' law:

$$\frac{dl_{cr}}{dt} = \frac{1}{t_0} \cdot \frac{dl_{cr}}{dN} = \frac{k_{cr}}{t_0}\left(\sigma_t b_{cr}\sqrt{\pi l_{cr}}\right)^{m_{cr}} \text{ for } \sigma_t > 0, \tag{18}$$

where $t_0$ is the time for one cycle, $b_{cr}$ is the stress intensity factor correction, $k_{cr}$ and $m_{cr}$ are constants that are determined from experimental data. The instantaneous rate of change of the crack area to volume ratio can be estimated by:

$$\frac{da_{cr}}{dt} = \frac{a_{\pm}\rho_{cr}w_{cr}}{t_0} \cdot \frac{dl_{cr}}{dt} = \frac{a_{\pm}\rho_{cr}w_{cr}}{t_0} \cdot k_{cr}\left(\sigma_t b_{cr}\sqrt{\pi l_{cr}}\right)^{m_{cr}} \text{ for } \sigma_t > 0, \tag{19}$$

For interactions between the SEI growth and particle cracking, we can apply the same SEI growth model on the cracks. However, the SEI formation on the newly exposed fresh crack surfaces will be faster than those surfaces with existing SEI layers. As a result, the SEI layer thickness is not uniform along cracks, because crack propagation leads to different exposure times for different interface locations along a crack. To avoid having different SEI thickness along cracks and simplify the problem, we have used the averaged thickness of SEI layer on cracks as the fundamental variable for the SEI on cracks sub-model, which is governed by:

$$\frac{\partial L_{SEI,cr}}{\partial t} = \frac{c_{int,Li}}{2L_{SEI,cr}} \cdot D_{int}\bar{V}_{SEI} \cdot e^{-(\phi_s - \phi_e)} - \frac{\partial l_{cr}}{\partial t} \cdot \frac{L_{SEI,cr}}{l_{cr}} \tag{20}$$

In the above equation, $L_{SEI,cr}$ changes for two reasons: (1) the existing SEI layers are growing (first term on the right), (2) the cracks expose fresh surfaces, which reduce the average SEI thickness on cracks (second term on the right).

*Mechanical LAM*

We consider loss of active materials (LAM) due to particle cracking here. The key equations are taken from Laresgoiti *et al.*[33] and Reniers *et al.*[34], and simplified by O'Kane *et al.*[11]:

$$\frac{\partial \varepsilon_a}{\partial t} = \frac{\beta}{t_0} \cdot \left(\frac{\sigma_h}{\sigma_c}\right)^{m_2} \text{ for } \sigma_h > 0, \tag{21}$$

among which $\varepsilon_a$ is the volume fraction of active materials, $\beta$ and $m_2$ are the LAM proportional and exponential terms, respectively. The hydrostatic stress $\sigma_h$ is a function of radial and tangential stress:

$$\sigma_h = (\sigma_r + 2\sigma_t)/3 \tag{22}$$

and $\sigma_c$ is the critical stress of the electrode materials.

*Solvent consumption*

The solvent consumption model is taken from Li *et al.* [25] The novelty of Li *et al.*'s [25] model is its flexibility to include the presence of an electrolyte reservoir, outside of the jelly roll but within the cell casing. Another novelty is that, to increase computational efficiency, Li *et al.* [25] did not use any differential equations, instead putting a wrapper around the main PyBaMM model to calculate how much solvent was consumed over a time interval $\Delta t$, then applying the effect of solvent consumption into the DFN model. In this work, as in Li *et al.*, [25] $\Delta t$ spans a set of ageing cycles and the solvent consumption is applied before each RPT cycle. Full details of the solvent consumption model, including further upgrades to allow it to interact with the degradation mechanisms in the *5 coupled* model, can be found in the Supplementary Information. However, in the *SEI + dry-out* model, it is assumed that SEI is the only other degradation mechanism, and that the reservoir is empty, in which case the equations are much simpler. The main effect of solvent consumption is to reduce the cross-sectional area $A_{cell}$:

$$\Delta n_{SEI} \equiv n_{SEI}(t + \Delta t) - n_{SEI}(t) = \frac{A_{cell}(t)}{\bar{V}_{SEI}} \int_0^{L_n} a_n[L_{SEI}(t + \Delta t) - L_{SEI}]dx \tag{23}$$

$$\Delta V_e \equiv V_e(t + \Delta t) - V_e(t) = \Delta n_{SEI}(\bar{V}_{SEI} - 2\bar{V}_{EC}) \tag{24}$$

$$R_{dry} \equiv \frac{V_e(t + \Delta t)}{V_e(t)} = \frac{V_e(t) + \Delta V_e}{V_e(t)} \tag{25}$$

$$V_e(t) = A_{cell}(t)[L_n\varepsilon_n(t) + L_s\varepsilon_s + L_p\varepsilon_p] \tag{26}$$

$$A_{cell}(t + \Delta t) = R_{dry}A_{cell}(t) \tag{27}$$

## *Ageing parameters*

The ageing parameters of this model is referred to O'Kane *et al.* [11] but tuned to fit the experimental data. To get a decent fit of the experiment data, some of the ageing parameters are first given a range, then we generate different combinations of these ageing parameters using Latin Hypercube sampling, the best fits are listed in Table 3. Those unchanged parameters are listed in Table S6 and Table S7.

Table 3 Ageing parameters varied during parameter sweeping and their best fits for the three models.

| Ageing mechanism | Parameter | Unit | SEI only | SEI + Dry out | 5 coupled |
|---|---|---|---|---|---|
| SEI | Inner SEI lithium interstitial diffusivity ($D_{\text{int}}$) | m²/s | 2.36e-18 | 1.2e-18 | 9.81e-19 |
| | Inner or outer SEI partial molar volume ($\bar{V}_{\text{SEI}}^{\text{inner}}$ or $\bar{V}_{\text{SEI}}^{\text{outer}}$) | m³/mol | 4e-5 | 6.74e-5 | 5.22e-5 |
| | SEI growth activation energy ($E_{\text{act}}^{\text{SEI}}$) | J/mol | 1e4 | 1e4 | 5e3 |
| Lithium plating | Dead lithium decay constant ($\gamma_0$) | 1/s | - | - | 1e-7 |
| | Lithium plating kinetic rate constant ($k_{\text{Li}}$) | m/s | - | - | 1e-10 |
| LAM model | Positive electrode LAM constant proportional term ($\beta_{\text{neg}}$) | 1/s | - | - | 2.98e-18 |
| | Negative electrode LAM constant proportional term ($\beta_{\text{pos}}$) | 1/s | - | - | 2.84e-9 |
| Mechanical and cracking | Negative electrode cracking rate ($k_{\text{cr}}^{\text{neg}}$) | | - | - | 5.29e-25 |
| DFN model | Negative electrode diffusivity activation energy ($E_{\text{act}}^{D_{s,n}}$) | J/mol | 6e4 | 6e4 | 2e4 |

## Acknowledgement

The authors thank Dr. Kieran O'Regan for providing updated open-circuit potential data for the LG M50 cell, using the same methods as his published work. The authors are also grateful for financial support from the EPSRC Faraday Institution Multiscale Modelling project (EP/S003053/1, grant number FIRG059). The first author is funded as a PhD student by the China Scholarship Council (CSC) Imperial Scholarship.

## Supplementary information

### Experiment details (RPT and ageing)

The commercial 21700 cylindrical cells (LGM50T, LG GBM50T2170) were cycled between 70%~85%SOC under 10, 25 and 40 °C. It has a $SiO_x$-doped graphite negative electrode alongside an NMC811 positive electrode, with a nominal 1C capacity of 5 Ah.

The overall test procedure includes a break-in test at the beginning of life (BOL), followed by repeated reference performance tests (RPT) and ageing tests. The break-in test is designed to bring different samples of a batch of cells into the same and stable conditions before the degradation study. It includes five full discharge−charge cycles at a rate of 0.2C. Two RPT tests are performed after the break-in test, namely a long one (last ~100 hours) and a short one (last ~50 hours), followed by an ageing test. After the ageing tests, the long RPT is performed after those even number of ageing tests, whereas the short RPT is performed after those odd number of ageing tests. The long RPT test contains 4 subsets, namely a full charge-discharge cycle at (i) 0.1C and (ii) 0.5C, and two galvanostatic intermittent titration technique (GITT) discharge tests at 0.5C with (iii) 25 and (iv) 5 pulses. For the 25-pulse GITT subset, 200 mAh of charge is passed, followed by a rest period of one hour. The 5-pulse GITT subset will pass 1000 mAh of charge, separated by the same rest period. To avoid overcharge/over-discharge, the lower and upper voltage limit for the full charge discharge cycle during the whole experiment are 2.5 V and 4.2 V, respectively.

For the ageing test, the cells are cycled between 70%~85%SOC, with the detailed steps listed in Table S1. During the whole experiment, the cells are fixed in a bespoke test rig which maintains a constant temperature on the base of the cell, with pseudo-adiabatic temperature conditions on the other surfaces.

The data extracted from the experiment for model validation in this paper are mainly from the RPT test, which includes the 0.1C discharge capacities, the 0.1C discharge voltage curves, the 0.1s resistance and the degradation modes (DMs) extracted from the 0.1C discharge voltage curves. Specifically, the 0.1s resistance is extracted from the 25-pulse GITT data based on the instantaneous potential drop upon applying the current pulse:

$$\text{Res} = \frac{V_2 - V_1}{I_2 - I_1} \tag{S1}$$

Among which $V_1$ and $I_1$ are the voltage and current during rest before applying the current pulse, and $V_2$ and $I_2$ are the values immediately after the current pulse. The term "immediately" here refers to 0.1 second here because the sampling rate of the GITT test is set to 10 Hz. Within such short period, the main contribution of the resistance should originate from the ohmic part (contact resistance, electrolyte conductivity, solid conductivity, etc.) and the charge transfer part, whereas the polarization part can be ignored. Note that using the above equations we can get 25 values of 0.1 resistance at different SOC. To compare the resistance changing over time and make it easier for model validation, we have picked the resistance at the 12$^{th}$ pulse, roughly corresponding to 52% SOC of the cells at BOL. This is because the resistance is relatively flat in this region. More information about the RPT test, test rig, thermal management, and data processing can be found in Kirkaldy et al.[26].

Table S1 Detailed cycling conditions during the aging test.

| Step | Control Type | Control Value | Primary Limits | Cell SOC after completion | Safety Limits |
|---|---|---|---|---|---|
| 1 | CC charge | 0.3C | $E_{cell}$ = 4.2 V | (100-y)% | $E_{cell}$ = 2.5 V |

| 2 | CV charge | 4.2 V | |I| < C/100 | 100% | N/A |
| 3 | Rest | Rest at OCV | time = 1 hour | N/A | N/A |
| 4 | CC discharge | 1C | Q = 730 mAh (=capacity*0.15) | 85% | $E_{cell}$ = 2.5 V |
| 5 | Rest | Rest at OCV | time = 3 hours | N/A | N/A |
| 6 | CC discharge | 1C | Q = 730 mAh (=capacity*0.15) | 70% | $E_{cell}$ = 2.5 V |
| 7 | CC charge | 0.3C | Q = 730 mAh (=capacity*0.15) | 85% | $E_{cell}$ = 4.2 V |
| 8 | Loop to step 6 | N/A | 516 times | N/A | N/A |

*The capacity used in these calculations is 4.865 Ah (from BOL characterisation)

**Modelling details**

*Solvent consumption*

The solvent consumption model is originated from Li et al [25], but has been upgraded to consider more complicated cases such as SEI on cracks, lithium plating, and the possibility of electrolyte being squeezed out. Therefore, it is necessary to introduce this model in detail here.

To start with, the solvent consumption model has the following key assumptions:

1. Fresh LIBs contain extra electrolyte outside the jelly roll but inside the cell package. This extra electrolyte is called electrolyte reservoir.
2. Solvate-volume effect is ignored. The solvent retains its initial volume before mixing in the composite electrolyte. $LiPF_6$ does not contribute to the volume of the electrolyte.
3. Only EC is consumed during cell ageing, which follows this reaction:

$$2Li^+ + 2(C_3H_4O_3)(EC) + 2e^- \leftrightarrow (CH_2OCO_2Li)_2 + C_2H_4, \quad (S2)$$

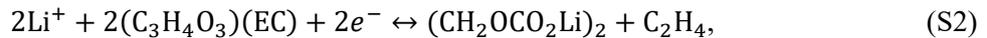

4. The time needed for the mass transport or electrolyte mixing between the reservoir and jelly roll is negligible.

5. Inside the jelly roll, EC is homogenous across the electrode, i.e., the EC concentration is independent of location.

To describe this extra electrolyte reservoir, three more variables are needed, the volume of the reservoir $V_e^{res}(t)$, the Li$^+$ and EC concentration of the reservoir, $c_{Li^+}^{res}(t)$ and $c_{EC}^{res}(t)$. Correspondingly, we use $A_{cell}(t)$, $c_{Li^+}^{JR}(x,t)$ and $c_{EC}^{JR}(t)$ to describe the electrolyte in the jelly roll.

The main idea of this model is to track these six variables. Therefore, the key is to determine how the porosity and solvent are changed due to all the side reaction. The porosity reduction due to SEI growth on both the negative particle surface and cracks as well as lithium plating (both reversible and irreversible) is:

$$\frac{d\varepsilon_n}{dt} = -\frac{d(L_{total} \cdot a_n)}{dt}, \tag{S3}$$

where $L_{total}$ is the total thickness of the deposit, including SEI and plated lithium:

$$L_{total} = L_{SEI} + 2l_{cr}w_{cr}\rho_{cr}L_{SEI,cr} + c_{Li} \cdot \frac{\bar{V}_{Li}}{a_n} + c_{dl} \cdot \frac{\bar{V}_{Li}}{a_n}. \tag{S4}$$

For simplicity, we assume the partial molar volume of dead lithium is the same as that of active lithium. We assume that the pore change of the cell only originates from the negative electrode. The total pore volume change of the cell is:

$$dV_{pore}^{JR} = d\int \varepsilon_n \cdot dV_{neg}, \tag{S5}$$

The solvent is consumed due to SEI growth (both on particle surfaces and cracks). Based on the SEI reaction in Eq. **Error! Reference source not found.**, the ratio of Li$^+$, solvent (we assume to be EC here), and SEI is 2:2:1 (the ratio of lithium moles to SEI moles, $z_{SEI} = 2$), therefore:

$$dn_{EC} = 2dn_{SEI} = 2 \cdot d\int -\frac{L_{SEI} + 2l_{cr}w_{cr}\rho_{cr}L_{SEI,cr}}{\bar{V}_{SEI}} \cdot a_n \cdot dV_{neg}, \tag{S6}$$

If the solvate-volume effect is ignored, the EC volume consumed will be:

$$dV_{EC} = dn_{EC} \cdot \bar{V}_{EC} = 2\bar{V}_{EC} \cdot d\int -\frac{L_{SEI} + 2l_{cr}w_{cr}\rho_{cr}L_{SEI,cr}}{\bar{V}_{SEI}} \cdot a_n \cdot dV_{neg}, \tag{S7}$$

To simplify the expression, we define a new symbol $B_{SEIcr}$:

$$B_{\text{SEIcr}}(L_{\text{SEI}}, L_{\text{SEI,cr}}) = 2\bar{V}_{\text{EC}} \cdot \frac{L_{\text{SEI}} + 2l_{\text{cr}}w_{\text{cr}}\rho_{\text{cr}}L_{\text{SEI,cr}}}{\bar{V}_{\text{SEI}}}, \tag{S8}$$

Then Eq. **Error! Reference source not found.**) becomes:

$$dV_{\text{EC}} = -d \int B_{\text{SEIcr}} \cdot a_n \cdot A_{\text{cell}} \cdot dL_{\text{neg}}, \tag{S9}$$

The difference between electrolyte volume reduction (due to EC consumption) and porosity reduction, is the original driving force of this solvent-consumption/dry-out model:

$$dV_e^{\text{need}} = dV_{\text{pore}}^{\text{JR}} - dV_{\text{EC}} = d\left(\int \varepsilon_n \cdot A_{\text{cell}} \cdot dL_{\text{neg}} + \int B_{\text{SEIcr}} \cdot a_n \cdot A_{\text{cell}} \cdot dL_{\text{neg}}\right) \tag{S10}$$

$dV_e^{\text{need}}$ is the electrolyte volume that the jelly roll "require" from the reservoir. However, the reservoir may not satisfy this requirement. We define $dV_e^{\text{add}}$ as the actual electrolyte volume move from reservoir to jelly roll. To make the derivation general enough to cope with different cases, we assume that from time $t$ to $t + dt$, electrolyte exchange occurs in both direction between the reservoir and jelly roll. Therefore, we further define $dV_e^{\text{squeeze}}$ as the electrolyte volume move from jelly roll to reservoir. The electrolyte volume changes of jelly roll and reservoir are:

$$dV_e^{\text{JR}} = dV_{\text{EC}} + dV_e^{\text{add}} - dV_e^{\text{squeeze}} \tag{S11}$$

$$dV_e^{\text{res}} = -dV_e^{\text{add}} + dV_e^{\text{squeeze}} \tag{S12}$$

Now based on the symbol of $dV_e^{\text{need}}$ and $V_e^{\text{res}}$, there are three cases to consider:

$$\begin{cases} dV_e^{\text{need}} > 0: dV_e^{\text{add}} = dV_e^{\text{need}} \cdot H(V_e^{\text{res}}); dV_e^{\text{squeeze}} = 0 \\ dV_e^{\text{need}} = 0: dV_e^{\text{add}} = 0, dV_e^{\text{squeeze}} = 0 \\ dV_e^{\text{need}} < 0: dV_e^{\text{add}} = 0; dV_e^{\text{squeeze}} = -dV_e^{\text{need}} \end{cases}, \tag{S13}$$

where:

$$H(x) = \begin{cases} 1, & x > 0 \\ 0, & x \leq 0 \end{cases}; H(-x) = \begin{cases} 1, & x < 0 \\ 0, & x \geq 0 \end{cases}, \tag{S14}$$

We further summarize the 3 cases in Eq. **Error! Reference source not found.**) to be:

$$dV_e^{\text{add}} = dV_e^{\text{need}} \cdot H(V_e^{\text{res}}) \cdot H(dV_e^{\text{need}}) \tag{S15}$$

$$dV_e^{\text{squeeze}} = -dV_e^{\text{need}} \cdot H(-dV_e^{\text{need}}) \tag{S16}$$

If we ignored the special case of re-wetting:

$$dV_e^{\text{JR}} = dV_{\text{EC}} + dV_e^{\text{need}} \cdot H(V_e^{\text{res}}) \cdot H(dV_e^{\text{need}}) + dV_e^{\text{need}} \cdot H(-dV_e^{\text{need}}) \tag{S17}$$

$$dV_e^{\text{res}} = -dV_e^{\text{need}} \cdot H(V_e^{\text{res}}) \cdot H(dV_e^{\text{need}}) - dV_e^{\text{need}} \cdot H(-dV_e^{\text{need}}) \tag{S18}$$

Recall that to describe the solvent consumption model, six variables are to tracked: $V_e^{\text{res}}(t)$, $c_{\text{Li}^+}^{\text{res}}(t)$, $c_{\text{EC}}^{\text{res}}(t)$, $A_{\text{cell}}(t)$, $c_{\text{Li}^+}^{\text{JR}}(x,t)$ and $c_{\text{EC}}^{\text{JR}}(t)$. $V_e^{\text{res}}(t)$ is tracked by Eq. **Error! Reference source not found.**). To track $A_{\text{cell}}(t)$, we define a dry-out ratio:

$$R_{\text{dry}} = \frac{V_e^{\text{JR}}(t+dt)}{V_{\text{pore}}^{\text{JR}}(t+dt)}, \tag{S19}$$

To do a bit more derivation, we have:

$$R_{\text{dry}} = \frac{V_e^{\text{JR}}(t)+dV_e^{\text{JR}}}{V_{\text{pore}}^{\text{JR}}(t)+dV_{\text{pore}}^{\text{JR}}} = \frac{V_e^{\text{JR}}(t)+dV_{\text{pore}}^{\text{JR}}-dV_e^{\text{need}}+dV_e^{\text{add}}}{V_e^{\text{JR}}(t)+dV_{\text{pore}}^{\text{JR}}} = 1 + \frac{dV_e^{\text{add}}-dV_e^{\text{need}}}{V_e^{\text{JR}}(t)+dV_{\text{pore}}^{\text{JR}}} = 1 + \frac{dV_e^{\text{need}} \cdot (H(V_e^{\text{res}}) \cdot H(dV_e^{\text{need}})-1)}{V_e^{\text{JR}}(t)+dV_{\text{pore}}^{\text{JR}}}, \tag{S20}$$

If $R_{\text{dry}} < 1$, dry-out occurs; if $R_{\text{dry}} > 1$, re-wetting occurs. However, one special case is when $R_{\text{dry}} > 1$ and $A_{\text{cell}}(t) = A_{\text{cell,max}}$, then the electrolyte area won't further increase, to include that, we define:

$$R_{\text{dry}'} = \begin{cases} 1 \text{ for } R_{\text{dry}} > 1 \text{ and } A_{\text{cell}}(t) = A_{\text{cell,max}} \\ R_{\text{dry}} \text{ for other conditions} \end{cases}, \tag{S21}$$

Then $A_{\text{cell}}(t)$ is now tracked by:

$$A_{\text{cell}}(t+dt) = R_{\text{dry}'} \cdot A_{\text{cell}}(t), \tag{S22}$$

Then the remaining four concentrations can be tracked by:

$$c_{\text{Li}^+}^{\text{res}}(t+dt) = \frac{n_{\text{Li}^+}^{\text{res}}(t+dt)}{V_e^{\text{res}}(t+dt)} = \frac{n_{\text{Li}^+}^{\text{res}}+dn_{\text{Li}^+}^{\text{res}}}{V_e^{\text{res}}+dV_e^{\text{res}}} = \frac{c_{\text{Li}^+}^{\text{res}}(t) \cdot (V_e^{\text{res}}-dV_e^{\text{add}})+dV_e^{\text{squeeze}} \cdot c_{\text{Li}^+}^{\text{JR,avg}}}{V_e^{\text{res}}-dV_e^{\text{add}}+dV_e^{\text{squeeze}}}, \tag{S23}$$

$$c_{\text{EC}}^{\text{res}}(t+dt) = \frac{n_{\text{EC}}^{\text{res}}(t+dt)}{V_e^{\text{res}}(t+dt)} = \frac{n_{\text{EC}}^{\text{res}}+dn_{\text{EC}}^{\text{res}}}{V_e^{\text{res}}+dV_e^{\text{res}}} = \frac{c_{\text{EC}}^{\text{res}}(t) \cdot (V_e^{\text{res}}-dV_e^{\text{add}})+dV_e^{\text{squeeze}} \cdot c_{\text{EC}}^{\text{JR}}}{V_e^{\text{res}}-dV_e^{\text{add}}+dV_e^{\text{squeeze}}}, \tag{S24}$$

$$c_{\text{Li}^+}^{\text{JR,avg}}(t+dt) = \frac{n_{\text{Li}^+}^{\text{JR}}(t+dt)}{V_e^{\text{JR}}(t+dt)} = \frac{n_{\text{Li}^+}^{\text{JR}} + dn_{\text{Li}^+}^{\text{JR}}}{V_e^{\text{JR}} + dV_e^{\text{JR}}} = \frac{c_{\text{Li}^+}^{\text{JR,avg}}(t) \cdot \left(V_e^{\text{JR}} - dV_e^{\text{squeeze}}\right) + c_{\text{Li}^+}^{\text{res}}(t) \cdot dV_e^{\text{add}}}{V_e^{\text{JR}} + dV_{\text{EC}} + dV_e^{\text{add}} - dV_e^{\text{squeeze}}}, \quad (S25)$$

$$c_{\text{EC}}^{\text{JR}}(t+dt) = \frac{n_{\text{EC}}^{\text{JR}}(t+dt)}{V_e^{\text{JR}}(t+dt)} = \frac{n_{\text{EC}}^{\text{JR}} + dn_{\text{EC}}^{\text{JR}}}{V_e^{\text{JR}} + dV_e^{\text{JR}}} == \frac{c_{\text{EC}}^{\text{JR,avg}}(t) \cdot \left(V_e^{\text{JR}} + dV_{\text{EC}} - dV_e^{\text{squeeze}}\right) + c_{\text{EC}}^{\text{res}}(t) \cdot dV_e^{\text{add}}}{V_e^{\text{JR}} + dV_{\text{EC}} + dV_e^{\text{add}} - dV_e^{\text{squeeze}}}, \quad (S26)$$

Notably, to get $c_{\text{Li}^+}^{\text{JR}}(x,t)$, we assume the added/removed electrolyte change the Li+ concentration with the same ratio:

$$c_{\text{Li}^+}^{\text{JR}}(x, t+dt) = \frac{c_{\text{Li}^+}^{\text{JR,avg}}(t+dt)}{c_{\text{Li}^+}^{\text{JR,avg}}(t)} \cdot c_{\text{Li}^+}^{\text{JR}}(x,t), \quad (S27)$$

*Doyle-Fuller-Newman (DFN) model and thermal model*

The form of equations for the DFN model and thermal model are taken from O'Regan *et al.* [27] and listed below.

Table S2 DFN model.

| Description | Equation | Boundary conditions |
|---|---|---|
| **Electrodes** | | |
| Mass conservation | $\frac{\partial c_{s,k}}{\partial t} = \frac{1}{r^2}\frac{\partial}{\partial r}\left(r^2 D_{s,k}\frac{\partial c_{s,k}}{\partial r}\right)$ | $\left.\frac{\partial c_{s,k}}{\partial r}\right|_{r=0} = 0,\ -D_{s,k}\left.\frac{\partial c_{s,k}}{\partial r}\right|_{r=R_k} = \frac{J_k}{a_k F}$ |
| Charge conservation | $\frac{\partial}{\partial x}\left(\sigma_{s,k}\frac{\partial \phi_{s,k}}{\partial x}\right) = J_k$ | $-\sigma_{s,n}\left.\frac{\partial \phi_{s,n}}{\partial x}\right|_{x=0} = -\sigma_{s,p}\left.\frac{\partial \phi_{s,p}}{\partial x}\right|_{x=L} = i_{app}$ <br> $-\sigma_{s,n}\left.\frac{\partial \phi_{s,n}}{\partial x}\right|_{x=L_n} = -\sigma_{s,p}\left.\frac{\partial \phi_{s,p}}{\partial x}\right|_{x=L-L_p} = 0$ |
| **Electrolyte** | | |
| Mass conservation | $\varepsilon_k\frac{\partial c_{e,k}}{\partial t} = \frac{\partial}{\partial x}\left(\varepsilon_k^b D_e\frac{\partial c_{e,k}}{\partial x}\right) + (1-t^+)\frac{J_k}{F}$ | $\left.\frac{\partial c_{e,n}}{\partial x}\right|_{x=0} = \left.\frac{\partial c_{e,p}}{\partial x}\right|_{x=L} = 0$ |
| Charge conservation | $\frac{\partial}{\partial x}\left(\varepsilon_k^b \sigma_{e,k}\left(\frac{\partial \phi_{e,k}}{\partial x} - \frac{2(1-t^+)RT}{F}\frac{\partial \log c_{e,k}}{\partial x}\right)\right) = -J_k$ | $\left.\frac{\partial \phi_{e,n}}{\partial x}\right|_{x=0} = \left.\frac{\partial \phi_{e,p}}{\partial x}\right|_{x=L} = 0$ |
| **Reaction kinetics** | | |
| Butler-Volmer | $J_k = \begin{cases} a_k j_{0,k}\ \sinh\left(\frac{1}{2}\frac{RT}{F}\eta_k\right), & k \in \{n,p\}, \\ 0, & k = s. \end{cases}$ | |
| Exchange current | $j_{0,k} = k_k\sqrt{c_{e,k} c_{s,k}(c_{s,k}^{\max} - c_{s,k})}\Big|_{r=R_k}$ | |
| Overpotential | $\eta_k = \phi_{s,k} - \phi_{e,k} - U_k\left(c_{s,k}\big|_{r=R_k}\right),\quad k \in \{n,p\}$ | |
| **Initial conditions** | | |
| Initial conditions | $c_{s,k} = c_{k0},\ c_{e,k} = c_{e0}$ | |
| **Terminal voltage** | | |
| Terminal voltage | $V = \phi_{e,p}\big|_{x=L} - \phi_{e,n}\big|_{x=0}$ | |

Table S3. Thermal model considered in this parameterization.

| Description | Equation |
|---|---|
| Energy conservation | $\rho C_p \frac{\partial T}{\partial t} = \nabla \cdot (k \nabla T) + q_{tot}$ |
| Total battery heat | $q_{tot} = q_{rev} + q_j + q_r$ |
| Reversible heat, $q_{rev}$ | $q_{rev} = \frac{\int_0^{l_n+l_s+l_p} j^{Li} T \left(\frac{\partial U}{\partial T}\right) dx}{l}$ |
| Entropy change | $\Delta S = nF \frac{\partial U}{\partial T}$ |
| Joule heat, $q_j$ | $q_j = \frac{\int_0^{l_n+l_s+l_p} \left[\sigma_{eff}\left(\frac{\partial \phi_s}{\partial x}\right)^2 + k_{eff}\left(\frac{\partial \phi_e}{\partial x}\right)^2 + \frac{2k_{eff}RT}{F}(1-t_+^0)\frac{\partial(\ln c_e)}{\partial x}\frac{\partial \phi_e}{\partial x}\right] dx}{l}$ |
| Reaction heat, $q_r$ | $q_r = \frac{\int_0^{l_n+l_s+l_p} j^{Li}(\phi_s - \phi_e - U) dx}{l}$ |
| Convection boundary condition | $-k \frac{\partial T}{\partial n} = h(T - T_{amb})$ |

*Zero-order hysteresis model*

PyBaMM has an optional zero-order hysteresis model, in which the either the lithiation or delithiation OCP is used, depending on the sign of the current:

$$U_p^{OCP} = \frac{1 + \tanh[100(J + 0.2)]}{2} U_{p,lith}^{OCP} + \frac{1 - \tanh[100(J + 0.2)]}{2} U_{p,delith}^{OCP} \qquad (S28)$$

$$U_n^{OCP} = \frac{1 - \tanh[100(J + 0.2)]}{2} U_{n,lith}^{OCP} + \frac{1 + \tanh[100(J + 0.2)]}{2} U_{n,delith}^{OCP} \qquad (S29)$$

where the applied current density J has units of A/m², is positive for discharge and negative for charge. The offset of 0.2 A/m2 is added so that the OCP corresponding to discharge is used when the cell is at rest. If it was not added, the hysteresis model would interfere with the GITT characterization because the OCP would change during the rest phases. The value of 0.2 A/m² is chosen because the magnitude of the current density curing the CV charge is always greater than this, so the hysteresis model will not interfere with the CV charge either.

**Input parameters**

This cell is chosen because it has been well parameterised for the DFN model by Chen *et al.*[28] and O'Regan *et al.*[27]. The availability of new open circuit potential (OCP) data enables us to make improvements to the electrode balancing, which are detailed below.

*Half-cell OCPs and electrode balancing*

New half-cell open circuit potential (OCP) data was provided by Dr. Kieran O'Regan, using the methods of Chen *et al.* The new dataset has two significant advantages over the published one. Firstly, the new data covers the 90-100% stoichiometry range in the graphite-silicon composite, which Chen *et al.* did not. Secondly, raw data for both discharge and charge is available, whereas only the raw data for discharge is included in the PyBaMM parameter set corresponding to Chen *et al.* The new and old OCPs are plotted in the top row of Fig. S1 for comparison.

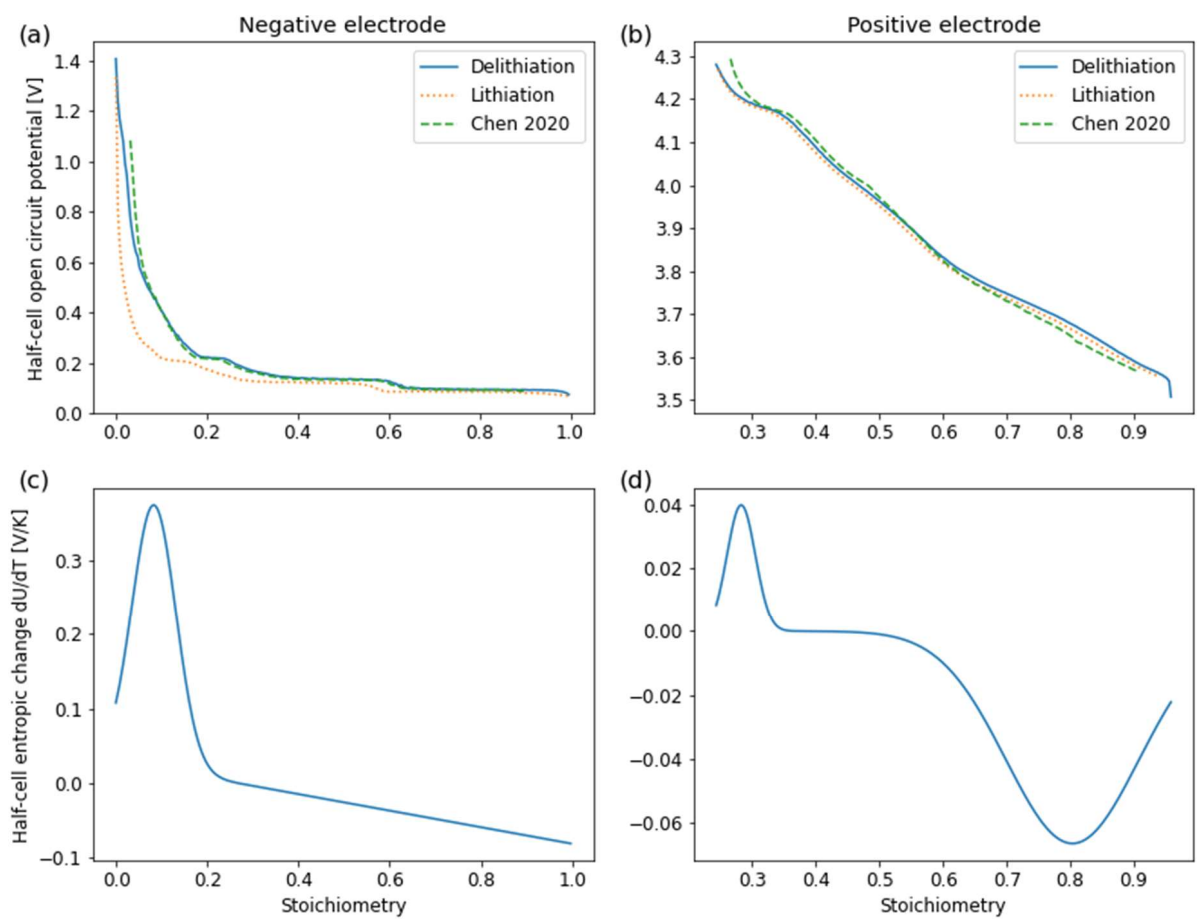

Fig. S1 Open-circuit potential and entropic change [27] of both electrodes.

Having raw data for both directions allows us to construct analytic OCP functions for both electrodes and both lithiation/delithiation directions:

$$U_{p,lith}^{OCV} = -0.7983 \cdot x + 4.513 - 0.03269 \cdot \tanh(19.83 \cdot (x - 0.5424)) - 18.23 \\ \cdot \tanh(14.33 \cdot (x - 0.2771)) + 18.05 \cdot \tanh(14.46 \cdot (x - 0.2776))$$ (S30)

$$U_{p,delith}^{OCV} = -0.7836 \cdot x + 4.513 - 0.03432 \cdot \tanh(19.83 \cdot (x - 0.5424)) - 19.35 \\ \cdot \tanh(14.33 \cdot (x - 0.2771)) + 19.17 \cdot \tanh(14.45 \cdot (x - 0.2776))$$ (S31)

$$U_{n,lith}^{OCV} = 0.5476 \cdot e^{-422.4 \cdot x} + 0.5705 \cdot e^{-36.89 \cdot x} + 0.1336 - 0.04758 \\ \cdot \tanh(13.88 \cdot (x - 0.2101)) - 0.01761 \cdot \tanh(36.2 \cdot (x - 0.5639)) \\ - 0.0169 \cdot \tanh(11.42 \cdot (x - 1))$$ (S32)

$$U_{n,delith}^{OCV} = 1.051 \cdot e^{-26.76 \cdot x} + 0.1916 - 0.05598 \cdot \tanh(35.62 \cdot (x - 0.1356)) - 0.04483 \\ \cdot \tanh(14.64 \cdot (x - 0.2861)) - 0.02097 \cdot \tanh(26.28 \cdot (x - 0.6183)) \\ - 0.02398 \cdot \tanh(38.1 \cdot (x - 1))$$ (S33)

where $x$ is the stoichiometry of the electrode.

Chen *et al.* did not provide any data for entropic changes of the two electrodes, but O'Regan *et al.* provided the following functions, which are plotted in the bottom row of Fig. S1:

$$\frac{\partial U_p}{\partial T} = 0.04006 \cdot e^{-\frac{(x-0.2828)^2}{0.0009855}} - 0.06656 \cdot e^{-\frac{(x-0.8032)^2}{0.02179}}$$ (S34)

$$\frac{\partial U_n}{\partial T} = -0.111 \cdot x + 0.02901 + 0.3562 \cdot e^{-\frac{(x-0.08308)^2}{0.004621}}$$ (S35)

The two advantages offered by the updated OCP data are significant because both the interstitial-diffusion limited SEI growth model and the lithium plating model are highly sensitive to the potential of the negative electrode surface, which means an accurate negative electrode OCP is essential to accurate prediction of degradation due to these mechanisms.

Chen *et al.* and O'Regan *et al.* disagree on the maximum and initial lithium concentrations in each electrode. Neither paper is consistent with the total capacity of the cell under (pseudo-)OCV conditions being 5 Ah; Chen *et al.*'s parameters result in a larger OCV capacity, while O'Regan *et al.*'s parameters result in a smaller one. We therefore conduct our own electrode balancing for the LG M50 cell.

Inspection of Fig. S1 (a) shows that the negative electrode OCP from O'Regan shows strong agreement with the one reported by Chen *et al.*, so we use the stoichiometry limits reported by Chen *et al.* (0.9014 and 0.0279) as a starting point. With all other parameters unchanged, a maximum lithium concentration of 32544 mol/m³ in the negative electrode results in a capacity of exactly 5 Ah between these limits under OCV conditions.

Upon inspection of Fig. S1 (b), the agreement between the positive electrode OCPs of Chen *et al.* and O'Regan is not as good. The shapes are very similar, but the data from Chen *et al.* is compressed into a smaller stoichiometry range than that from O'Regan *et al.* To rectify this, new stoichiometry limits of 0.9256 and 0.2411 are calculated using a bisection method and Eq. (S30) to find the stoichiometries that result in the same OCPs as the limits from Chen *et al.* (0.9084 and 0.27) did for the OCP function in that paper. With all other parameters unchanged, a maximum lithium concentration of 32544 mol/m³ in the negative electrode results in a (pseudo-)OCV capacity of exactly 5 Ah.

However, Kirkaldy *et al.* measured a smaller capacity of 4.865 Ah. We assume this is due to the cells having degraded due to SEI formation while the cells were in storage. The upper stoichiometry limit is reduced to 0.8771, to account for the difference in capacity.

The SEI thickness, SEI on cracks thickness and negative porosity at BOL are changed accordingly, see **Error! Reference source not found.**. We assume that the SEI is homogenously distributed on the negative electrode particle surface and crack surface, each has two layers with the same thickness:

$$L_{\text{SEI},0}^{\text{inner}} = L_{\text{SEI},0}^{\text{outer}} = L_{\text{SEIcr},0}^{\text{inner}} = L_{\text{SEIcr},0}^{\text{outer}} = 1800 \cdot \frac{Q_{\text{loss},0}}{F \cdot z_{\text{SEI}}} \cdot \frac{\bar{V}_{\text{SEI}}}{S_r a_n V}, \quad (S36)$$

among which $Q_{\text{loss},0}$ is the initial capacity loss (5-4.865=0.135Ah), F is the Faraday constant, $z_{\text{SEI}}$ is the ratio of lithium moles to SEI moles during the SEI reaction. $\bar{V}_{\text{SEI}}$ is the SEI partial molar volume (in $\frac{m^3}{mol}$), $S_r$ is the roughness ratio, which can be understood as a coefficient to convert the "normal actual surface area of a particle without cracks" to "total area that can growth SEI" when cracks are presented:

$$S_r = 2l_{cr}w_{cr}\rho_{cr} + 1, \tag{S37}$$

where $l_{cr}, w_{cr}, \rho_{cr}$ are crack length, crack width and number of cracks per unit area, respectively.

$a_n$ is the surface aspect ratio, defined as:

$$a_n = \frac{3\varepsilon_{s,n}}{R_n}, \tag{S38}$$

where $\varepsilon_{s,n}$ and $R_n$ are the active material volume fraction and particle radius of the negative electrode. $V$ is the volume of the negative electrode (pores included).

Porosity reduction due to such SEI growth is:

$$\Delta\varepsilon = 2L_{SEI,0}^{inner} \cdot S_r a = 3600 \cdot \frac{Q_{loss,0}}{F \cdot z_{SEI}} \cdot \frac{\bar{V}_{SEI}}{V} \tag{S39}$$

The initial negative porosity from O'Regan *et al.* for a fresh LG M50 cell with a capacity of 5 Ah is 0.25. In our study, the actual initial negative porosity is:

$$\varepsilon_{n,0} = 0.25 - \Delta\varepsilon \tag{S40}$$

In our study, we have made $\bar{V}_{SEI}$ a tuning parameter, which will change both the initial negative electrode porosity and the four SEI thicknesses in Eq. (S36). As an example, for $\bar{V}_{SEI} = 9.585 \cdot 10^{-5} \frac{m^3}{mol}$, the four initial SEI thicknesses and initial negative electrode porosity will be $1.236 \cdot 10^{-8}$ m and 0.222, respectively.

*Electrolyte parameters*

The electrolyte diffusivity, conductivity, cation transference number and thermodynamic factor (Fig. S1) are based on the EC:EMC 3:7 wt% in LiPF$_6$ from Landesfeind and Gasteiger[35]. We manually add the saturation limit of 4000 mol/m³ (4M), assuming that at any salt concentration higher than this value, salt precipitation will happen, and the four properties will behave as if the concentration is 4000 mol/m3 for such time until it drops below this value and the salt dissolves back into the solvent. Coping with the complex precipitation / dissolution dynamics is beyond the scope of this work.

$$c_e^{cor} = \begin{cases} c_e/1000, & c_e < 4000 \\ 4, & c_e \geq 4000 \end{cases} \tag{S41}$$

$$D_\text{e} = 10^{-10} \cdot 1010 \cdot e^{1.01 \cdot c_\text{e}^\text{cor}} \cdot e^{-1560/T} \cdot e^{c_\text{e}^\text{cor} \cdot (-487)/T} \tag{S42}$$

$$\kappa_\text{e} = 0.1 \cdot 0.521 \cdot (1 + (T - 228)) \cdot c_\text{e}^\text{cor} \cdot \frac{\left(1 - 1.06 \cdot \sqrt{c_\text{e}^\text{cor}} + 0.8353 \cdot \left(1 - 0.00359 \cdot e^{\frac{1000}{T}}\right) \cdot c_\text{e}^\text{cor}\right)}{1 + (c_\text{e}^\text{cor})^4 \cdot \left(0.00148 \cdot e^{\frac{1000}{T}}\right)} \tag{S43}$$

$$\begin{aligned}t_+^0 = &-12.8 - 0.612 \cdot c_\text{e}^\text{cor} + 0.0821 \cdot T + 0.904 \cdot (c_\text{e}^\text{cor})^2 + 0.0318 \cdot c_\text{e}^\text{cor} \cdot T - 1.27 \cdot 10^{-4} \\ &\cdot T^2 + 0.0175 \cdot (c_\text{e}^\text{cor})^3 - 3.12 \cdot 10^{-3} \cdot (c_\text{e}^\text{cor})^2 \cdot T - 3.96 \cdot 10^{-5} \cdot c_\text{e}^\text{cor} \cdot T^2\end{aligned} \tag{S44}$$

$$\begin{aligned}\chi = &\, 25.7 - 45.1 \cdot c_\text{e}^\text{cor} - 0.177 \cdot T + 1.94 \cdot (c_\text{e}^\text{cor})^2 + 0.295 \cdot c_\text{e}^\text{cor} \cdot T + 3.08 \cdot 10^{-4} \cdot T^2 \\ &+ 0.259 \cdot (c_\text{e}^\text{cor})^3 - 9.46 \cdot 10^{-3} \cdot (c_\text{e}^\text{cor})^2 \cdot T - 4.54 \cdot 10^{-4} \cdot c_\text{e}^\text{cor} \cdot T^2\end{aligned} \tag{S45}$$

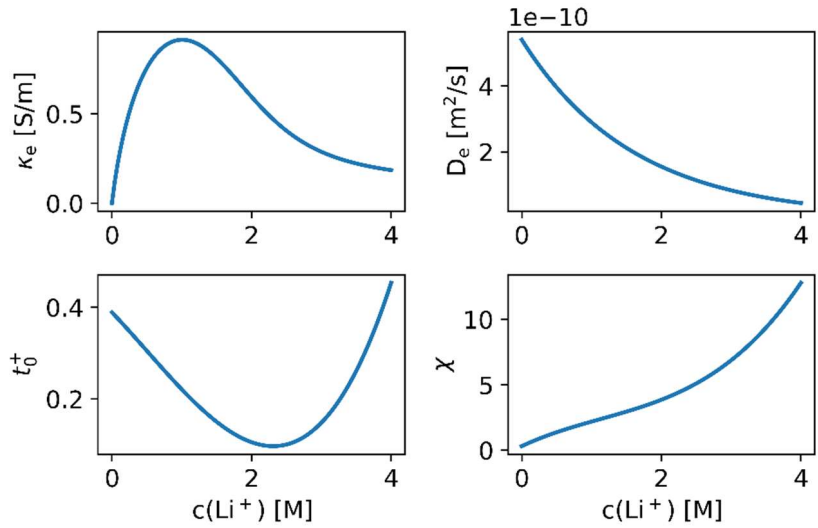

Fig. S1 Electrolyte conductivity, diffusivity, cation transference number and thermodynamic factor for EC:EMC 3:7 wt% in LiPF$_6$[35] at 25 °C.

## Other parameters

The remaining parameters come from O'Regan et al.[27] and are unchanged. All beginning of life parameters are listed in Tables S4 and S5, where the parameters that have been changed for this work are indicated with an asterisk (*).

Table S4. The parameters used for the DFN and thermal model in this study. All parameters are taken from O'Regan et al. except for those marked with an asterisk (*), which have been changed for this work.

| Type | Parameter | Unit | Positive electrode | Separator | Negative electrode |
| --- | --- | --- | --- | --- | --- |

|  | Active material |  | $Li_xNi_{0.8}Mn_{0.1}Co_{0.1}O_2$ | Ceramic coated polyolefin | $Li_xC_6 + SiO_x$ |
|---|---|---|---|---|---|
| Design specifications | Current collector thickness ($L_{CC}$) | M | $1.6 \cdot 10^{-5}$ |  | $1.2 \cdot 10^{-5}$ |
|  | Current collector conductivity ($\sigma_{CC}$) | S/m | $3.6914 \cdot 10^7$ |  | $5.8411 \cdot 10^7$ |
|  | Current collector density ($\rho_{CC}$) | kg/m³ | 2700 |  | 8960 |
|  | Current collector specific heat capacity ($C_{p,cc}$) | J/kg/K | 897 |  | 385 |
|  | Current collector thermal conductivity ($\lambda_{cc}$) | W/m/K | 237 |  | 401 |
|  | Electrode thickness ($L$) | M | $7.56 \cdot 10^{-5}$ | $1.2 \cdot 10^{-5}$ | $8.52 \cdot 10^{-5}$ |
|  | Electrode length ($w$) | M |  | 1.58 |  |
|  | Electrode width ($h$) | M |  | $6.5 \cdot 10^{-2}$ |  |
|  | Cell cooling surface area ($A_{cool}$) | m² |  | $5.31 \cdot 10^{-3}$ |  |
|  | Cell volume ($V_{cell}$) | m³ |  | $2.42 \cdot 10^{-5}$ |  |
|  | Cell thermal expansion coefficient ($\alpha_{th}$) | m/K |  | $1.1 \cdot 10^{-6}$ |  |
|  | Total heat transfer coefficient ($h_{th}$) | W/m²/K |  | 20 |  |
|  | Mean particle radius ($R_s$) | M | $5.22 \cdot 10^{-6}$ |  | $5.86 \cdot 10^{-6}$ |
|  | Electrolyte volume fraction ($\varepsilon_e$) |  | 0.335 | 0.47 | See Eq. (S40)* |
|  | Active material volume fraction ($\varepsilon_s$) |  | 0.665 |  | 0.75 |
|  | Contact resistance ($R_0$) | mΩ |  | 11.5 |  |
|  | Bruggeman exponent (electrode) ($b$) |  | 0 | 1.5 | 0 |
|  | Bruggeman exponent (electrolyte) ($b$) |  | 1.5 | 1.5 | 1.5 |
| Electrode | Solid phase lithium diffusivity ($D_s$) | m²·s⁻¹ | Eq. (S46) |  | Eq. (S47) |
|  | Solid phase electronic conductivity ($\sigma_s$) | S·m⁻¹ | Eq. (S48) |  | 215 |
|  | Density (wet, $\rho_{s+l}$) | kg/m³ | 3700 | 1548 | 2060 |
|  | Density (porous, $\rho_s$) | kg/m³ | 3270 | 1740 | 946 |
|  | Poisson's ratio ($\nu$) |  | 0.2 |  | 0.3 |
|  | Young's modulus ($E$) | GPa | 375 |  | 15 |
|  | Reference concentration for free of deformation ($c_{ref}$) | mol·m⁻³ | 0 |  | 0 |
| | Partial molar volume ($\Omega$) | m³/mol | 1.25·10⁻⁵ | | 3.1·10⁻⁶ |
| | Wet electrode specific heat capacity ($\bar{\theta}$) | J/kg/K-1 | Eq. (S49) | Eq. (S49) | Eq. (S49) |
| | Thermal conductivity (wet, $\lambda$) | W/m/K | Eq. (S53) | 0.3344 | Eq. (S54) |
| | Maximum concentration ($c_s^{max}$) | mol·m⁻³ | 52787* | | 32544* |
| | Initial concentration ($c_s^{int}$) | mol·m⁻³ | 12727* | | 28543* |
| Electrolyte | Li⁺ diffusivity in the electrolyte ($D_e$) | m²·s⁻¹ | | EC:EMC 3:7 wt% in LiPF₆, Eq. (S42)* | |
| | Electrolyte ionic conductivity ($\kappa_e$) | S·m⁻¹ | | EC:EMC 3:7 wt% in LiPF₆, Eq. (S43)* | |
| | Cation transference number ($t_+^0$) | - | | EC:EMC 3:7 wt% in LiPF₆, Eq. (S44)* | |
| | Thermodynamic factor ($\chi$) | - | | EC:EMC 3:7 wt% in LiPF₆, Eq. (S45)* | |
| | Initial Li⁺ concentration in electrolyte in the jell roll ($c_{Li^+,0}^{JR}$) | mol·m⁻³ | | 1000 | |
| | Initial EC concentration in electrolyte in the jell roll ($c_{EC,0}^{JR}$) | mol·m⁻³ | | 4541 | |
| | Heat capacity of electrolyte ($C_{p,l}$) | J/K/kg³ | | 229 | |
| | ~~~ | ``` | ```` | ````` | |
| | ```` | ```` | | ```` | |
| Intercalation reaction | Open Circuit Voltages ($U_i^{OCV}$) | V | Eq. (S30), (S31)* | | Eq. (S32), (S33)* |
| | Entropy change ($\frac{\partial U_i}{\partial T}$) | V/K | Eq. (S34) | | Eq. (S35) |
| | Exchange current density ($j_{0,k}^{int}$) | A/m² | Eq. (S55) | | Eq. (S56) |

Solid-phase diffusivity for negative and positive electrode (Fig. S3) follows:

$$\log_{10}(D_s^{ref}/R_{cor}) = a_0 \cdot x + b_0 + a_1 \cdot e^{-\frac{(x-b_1)^2}{c_1}} + a_2 \cdot e^{-\frac{(x-b_2)^2}{c_2}} + +a_3 \cdot e^{-\frac{(x-b_3)^2}{c_3}} + a_4 \cdot e^{-\frac{(x-b_4)^2}{c_4}} \quad (S46)$$

$$D_s = D_s^{ref} \cdot e^{-\frac{E_{act}}{R}\left(\frac{1}{T} - \frac{1}{T_{ref}}\right)} \quad (S47)$$

Table S5. Fitting parameters for the function describing solid-phase electrode diffusivity[27]. A "-" means that the term including that parameter has not been included. The activation energy has been changed for this work.

| Fitting Parameter | Positive Electrode | Negative Electrode |
| --- | --- | --- |
| $a_0$ | - | 11.17 |
| $a_1$ | -0.9231 | -1.553 |
| $a_2$ | -0.4066 | -6.136 |
| $a_3$ | -0.993 | -9.725 |
| $a_4$ | - | 1.85 |
| $b_0$ | -13.96 | -15.11 |
| $b_1$ | 0.3216 | 0.2031 |
| $b_2$ | 0.4532 | 0.5375 |
| $b_3$ | 0.8098 | 0.9144 |
| $b_4$ | - | 0.5953 |
| $c_1$ | 0.002534 | 0.0006091 |
| $c_2$ | 0.003926 | 0.06438 |
| $c_3$ | 0.09924 | 0.0578 |
| $c_4$ | - | 0.001356 |
| $E_{act}$ | 12000 | See Table S6 in Methods section* |
| $R_{cor}$ | 2.7 | 3.0321 |

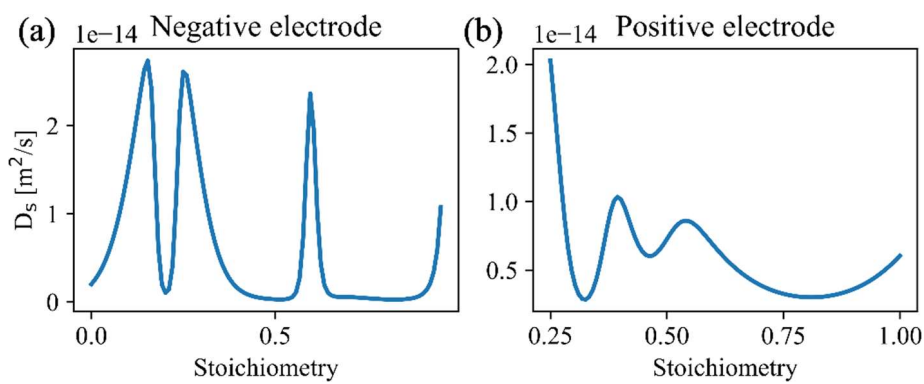

Fig. S3 Solid-phase diffusivity of both electrodes[27] at 25 °C.

The positive electrode electronic conductivity is:

$$\sigma_{s,p} = 0.8473 \cdot e^{-\frac{3500}{R} \cdot \left(\frac{1}{T} - \frac{1}{298.15}\right)} \tag{S48}$$

The specific heat capacity of wet electrode $\bar{\theta}$ is calculated by combining those of each bulk materials in the electrode:

$$\bar{\theta} = \rho_s \cdot C_{p,s} \cdot \varepsilon_s + \rho_l \cdot C_{p,l} \cdot \varepsilon_l, \tag{S49}$$

where $\rho$, $C_p$, and $\varepsilon$ are the density, gravimetric heat capacity and volume fraction for the porous electrode/separator and electrolyte, respectively. The specific heat capacities of the solid parts (electrode and separator) are:

$$C_{p,pos} = -0.0008414 \cdot T^3 + 0.7892 \cdot T^2 - 241.3 \cdot T + 2.508 \cdot 10^4 \tag{S50}$$

$$C_{p,neg} = 0.0004932 \cdot T^3 - 0.491 \cdot T^2 + 169.4 \cdot T - 1.897 \cdot 10^4 \tag{S51}$$

$$C_{p,sep} = 0.001494 \cdot T^3 - 1.444 \cdot T^2 + 475.5 \cdot T - 5.13 \cdot 10^4 \tag{S52}$$

Note $T$ in the above equations is in Kelvin.

The thermal conductivities of wet electrodes are:

$$\lambda_{pos} = 2.063 \cdot 10^{-5} \cdot T^2 - 0.01127 \cdot T + 2.331 \tag{S53}$$

$$\lambda_{neg} = -2.61 \cdot 10^{-4} \cdot T^2 + 0.1726 \cdot T - 24.49 \tag{S54}$$

Exchange current densities for intercalation of both electrodes are:

$$j_{0,p}^{int} = 5.028 \cdot e^{-\frac{2.401 \cdot 10^4}{R}\left(\frac{1}{T}-\frac{1}{298.15}\right)} \cdot \left(\frac{c_e}{\langle c_e \rangle}\right)^{0.57} \cdot \left(\frac{c_{s,suf}}{c_{s,max}}\right)^{0.43} \cdot \left(1 - \frac{c_{s,suf}}{c_{s,max}}\right)^{0.57} \tag{S55}$$

$$j_{0,n}^{int} = 2.668 \cdot e^{-\frac{4 \cdot 10^4}{R}\left(\frac{1}{T}-\frac{1}{298.15}\right)} \cdot \left(\frac{c_e}{\langle c_e \rangle}\right)^{0.208} \cdot \left(\frac{c_{s,suf}}{c_{s,max}}\right)^{0.792} \cdot \left(1 - \frac{c_{s,suf}}{c_{s,max}}\right)^{0.208} \tag{S56}$$

**Unchanged degradation parameters**

Some of the degradation parameters are tuned to give the best fit to experimental data; these are listed in Table 3 in the Methods section. The remaining degradation parameters are unchanged from the values used by O'Kane *et al.*, except for the solvent consumption parameters, which are unchanged from the values used by Li *et al.*.

Table S6 Ageing parameters (related to the negative electrode only) that remain unchanged in this study.

| Ageing mechanism | Parameter | Unit | Values |
|---|---|---|---|
| SEI | Ratio of lithium moles to SEI moles ($z_{SEI}$) | | 2 |
| | Lithium interstitial reference concentration ($c_{int,Li}$) | mol/m$^3$ | 15 |
| | SEI resistivity ($\rho_{SEI}$) | $\Omega \cdot m$ | 2e5 |
| | Inner SEI reaction proportion ($\alpha$) | - | 0.5 |
| | Initial inner SEI thickness ($L_{SEI,inner,0}$) | m | 1.23625e-08 |
| | Initial outer SEI thickness ($L_{SEI,outer,0}$) | m | 1.23625e-08 |
| | Initial Li$^+$ concentration in electrolyte in the reservoir ($c_{Li^+,0}^{res}$) | mol/m$^3$ | 1000 |
| | Initial EC concentration in electrolyte in the reservoir ($c_{EC,0}^{res}$) | mol/m$^3$ | 4541 |
| Solvent consumption | Initial Li$^+$ concentration in electrolyte in the reservoir ($c_{Li^+,0}^{res}$) | mol/m$^3$ | 1000 |
| | Initial excessive electrolyte amount | - | 1.0 |
| | Initial EC concentration in electrolyte in the reservoir ($c_{EC,0}^{res}$) | mol/m$^3$ | 4541 |
| | EC partial molar volume ($\bar{V}_{EC}$) | m$^3$/mol | 6.667e-5 |
| Lithium plating | Lithium plating transfer coefficient ($\alpha_{a,Li}$) | - | 0.65 |
| | Initial plated lithium concentration ($c_{Li,0}$) | mol/m$^3$ | 0 |
| | Lithium metal partial molar volume ($\bar{V}_{Li}$) | m$^3$/mol | 1.3e-5 |

Table S7 Ageing parameters (related to both electrodes) that remain unchanged in this study.

| Ageing mechanism | Parameter | Unit | Positive | Negative |
|---|---|---|---|---|
| LAM model | LAM exponential term ($m_2$) | - | 2 | 2 |
| Mechanical and cracking | Electrode stress intensity factor correction ($b_{cr}$) | | 1.12 | 1.12 |
| | Paris' law exponential term ($m_{cr}$) | | 2.2 | 2.2 |
| | Number of cracks per unit area ($\rho_{cr}^{neg}$) | 1/m$^2$ | 3.18e15 | 3.18e15 |

| | | | |
|---|---|---|---|
| Initial crack length ($l_{cr,0}^{pos}$) | m | $2 \cdot 10^{-8}$ | $2 \cdot 10^{-8}$ |
| Initial crack width ($w_{cr,0}^{pos}$) | m | $1.5 \cdot 10^{-8}$ | $1.5 \cdot 10^{-8}$ |
| Electrode critical stress ($\sigma_c$) | MPa | 375 | 60 |

**Other figures**

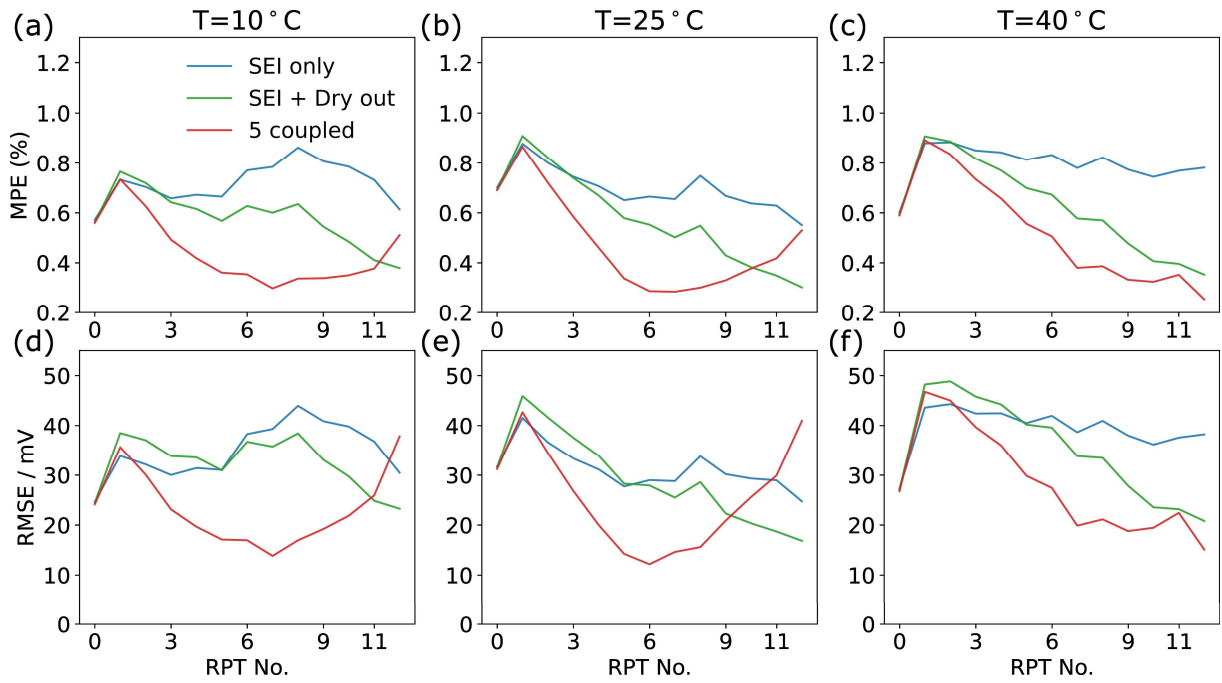

Fig. S2 MPE (a ~ c) and RMSE (d ~ e) of voltage.

Table S8. Mean RMSE of voltage fitting for all C/10 discharge RPT cycles.

| Mean RMSE / mV | 10°C | 25°C | 40°C |
|---|---|---|---|
| *SEI only* | 34.83 | 31.31 | 39.39 |
| *SEI + Dry out* | 32.33 | 29.16 | 35.13 |
| *5 coupled* | 23.26 | 25.33 | 28.34 |

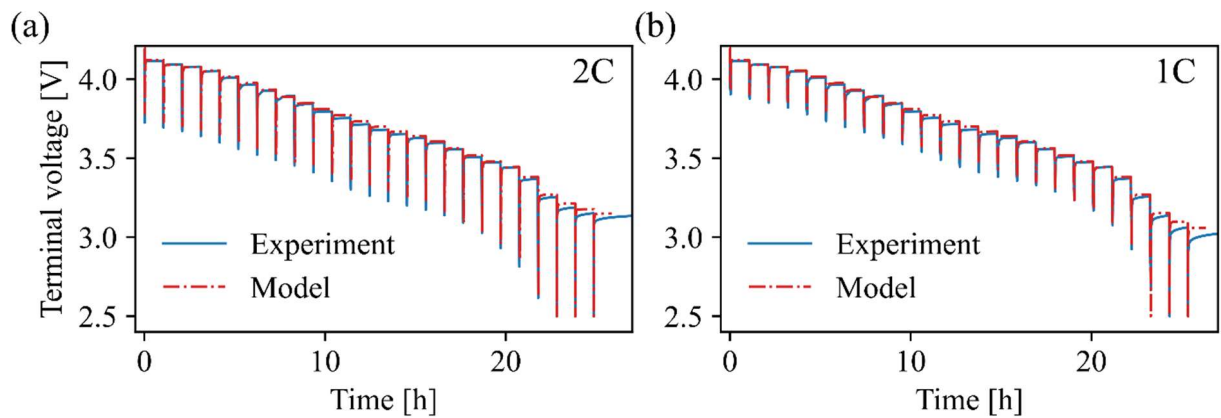

Fig. S3 Validation against GITT (1C and 2C) at BOL (from double-solvent paper)

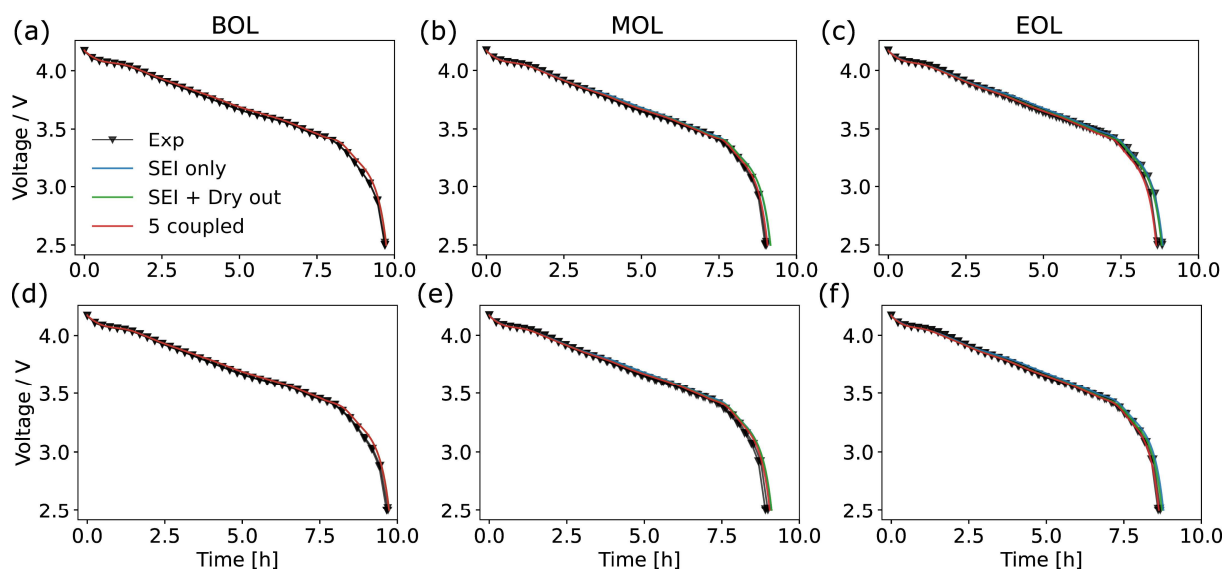

Fig. S4 Validation of the voltage at 10 °C (a ~ c) and 40 °C (d ~ f).

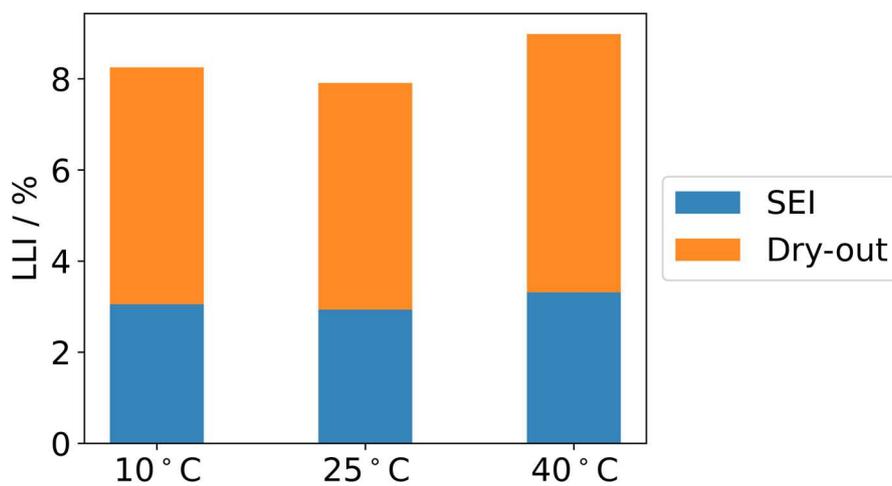

Fig. S5 Contribution of SEI and Dry-out to LLI in the *SEI + Dry out* model (diffusion slow due to activation energy)

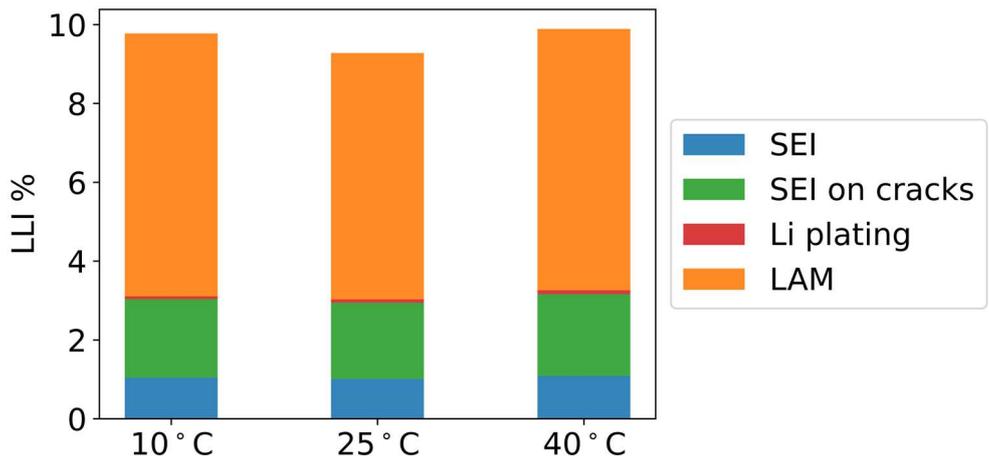

Fig. S6 Different contributions to LLI in the *5 coupled* model

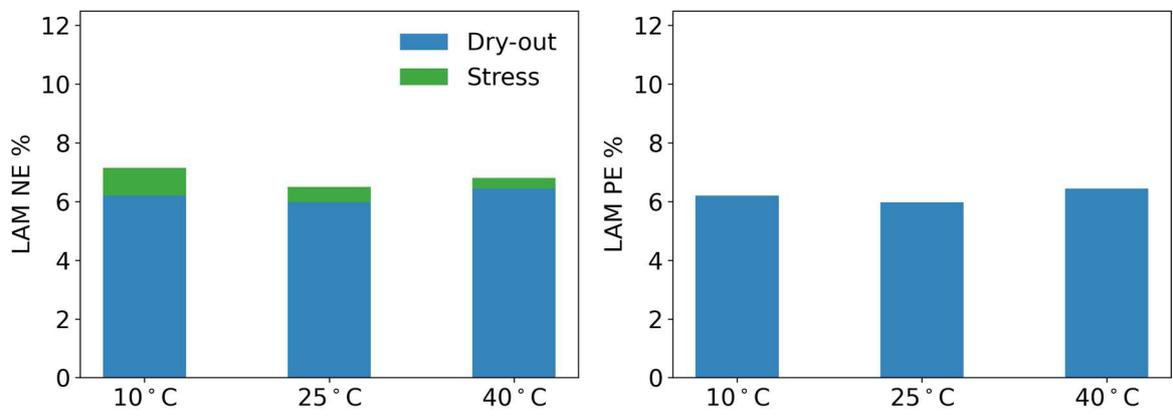

Fig. S7 Different contributions to LAM in the *5 coupled* model

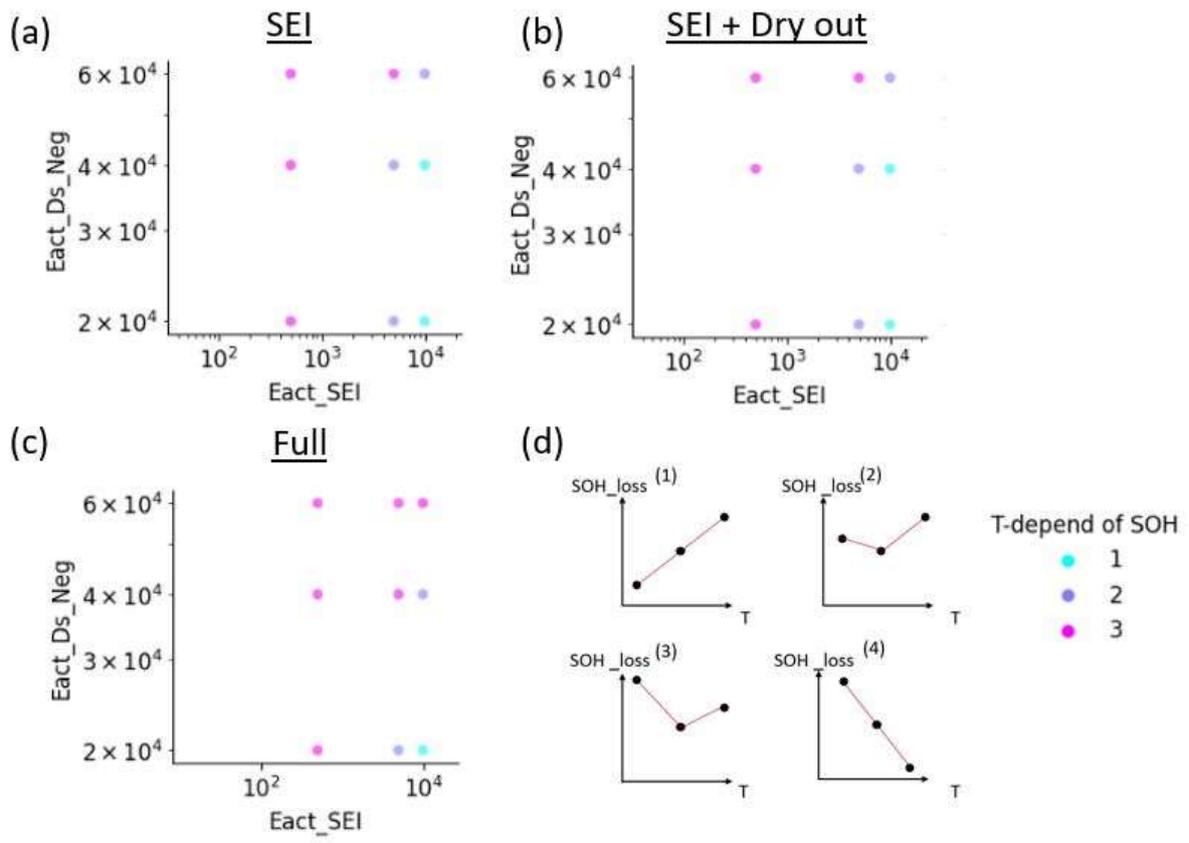

Fig. S8 T dependency of $E_{\text{act}}^{\text{SEI}}$ and $E_{\text{act}}^{\text{Ds,n}}$.

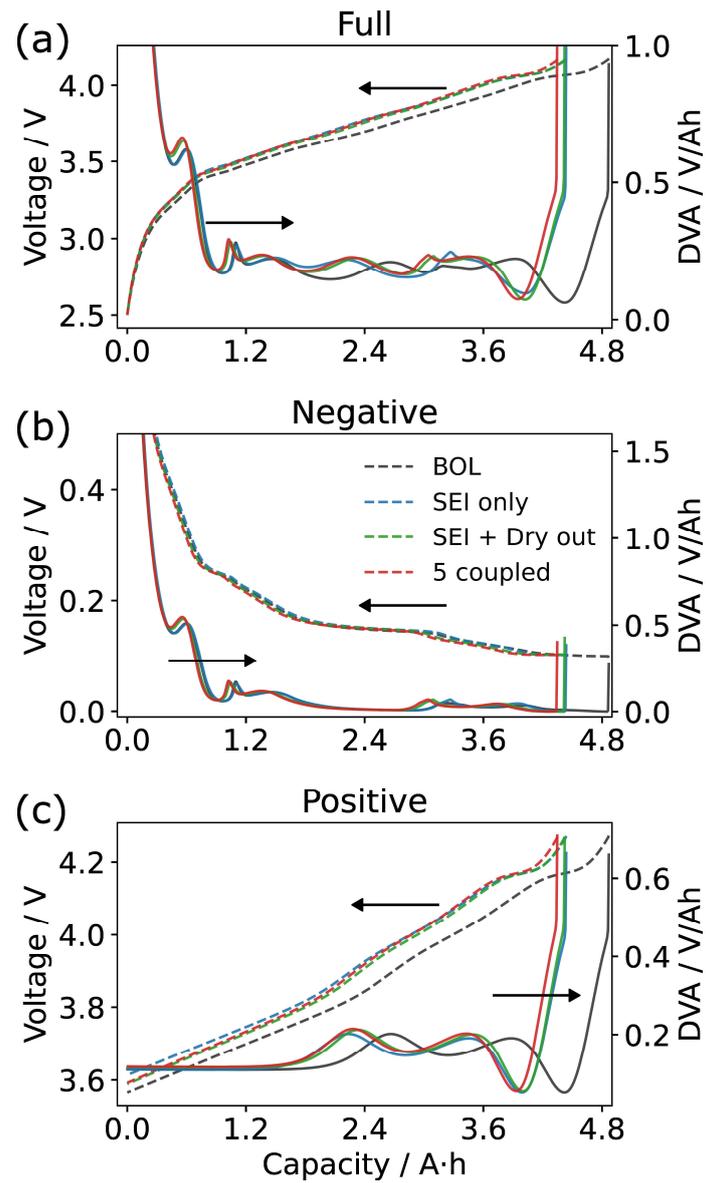

Fig. S9 Effect on half-cell potentials (dashed lines) and DVA (solid lines) during C/10 charge of the cell aged at 25 °C (BOL vs. EOL-RPT-13).

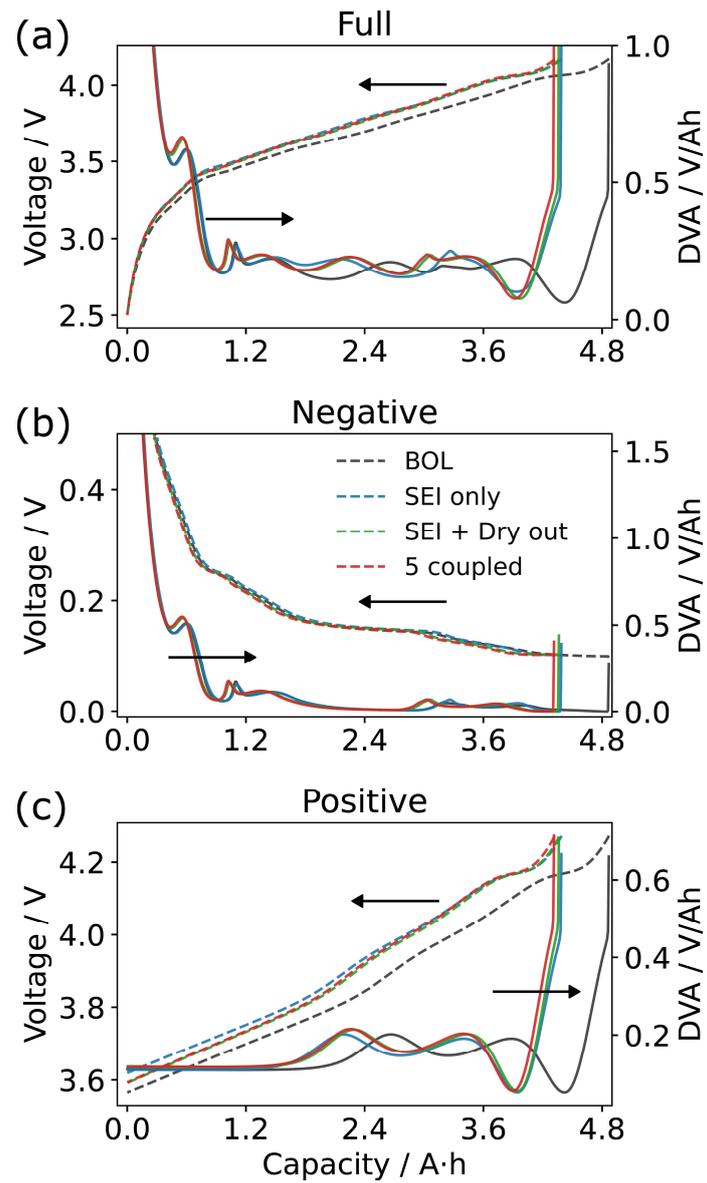

Fig. S10 Effect on half-cell potentials (dashed lines) and DVA (solid lines) during C/10 charge of the cell aged at 40 °C (BOL vs. EOL-RPT-13).